\documentclass[sigconf]{acmart}

\AtBeginDocument{%
  }

\copyrightyear{2025}
\acmYear{2025}
\setcopyright{acmlicensed}
\acmConference[KDD '25]{Proceedings of the 31st ACM
SIGKDD Conference on Knowledge Discovery and Data Mining V.2}{August 3--7,
2025}{Toronto, ON, Canada}
\acmBooktitle{Proceedings of the 31st ACM SIGKDD Conference on Knowledge
Discovery and Data Mining V.2 (KDD '25), August 3--7, 2025, Toronto, ON, Canada}
\acmDOI{10.1145/3711896.3737036}
\acmISBN{979-8-4007-1454-2/2025/08}

\usepackage{enumitem}
\usepackage{multirow}
\usepackage{makecell}
\usepackage{threeparttable}
\usepackage{tcolorbox}
\usepackage{color, xspace}
\usepackage{pifont}
\begin{document}

\title{LSRP: A Leader–Subordinate Retrieval Framework for Privacy-Preserving Cloud–Device Collaboration}

\author{Yingyi Zhang}
\authornote{Contributed equally to this work.}
\orcid{0000-0001-9062-3428}
\affiliation{
  \institution{Dalian University of Technology \& City University of Hong Kong}
  \city{Dalian}
  \country{China}
}
\email{yingyizhang@mail.dlut.edu.cn}

\author{Pengyue Jia}
\authornotemark[1]
\orcid{0000-0003-4712-3676}
\affiliation{
  \institution{City University of Hong Kong}
  \city{Hong Kong}
  \country{Hong Kong SAR}
}
\email{jia.pengyue@my.cityu.edu.hk}

\author{Xianneng Li}
\authornote{Corresponding authors.}
\orcid{0000-0003-4130-6930}
\affiliation{
  \institution{Dalian University of Technology}
  \city{Dalian}
  \country{China}
}
\email{xianneng@dlut.edu.cn}

\author{Derong Xu}
\orcid{0000-0002-3971-9907}
\affiliation{
  \institution{University of Science and Technology of China \& City University of Hong Kong}
  \city{Hefei}
  \country{China}
}
\email{derongxu@mail.ustc.edu.cn}

\author{Maolin Wang}
\orcid{0000-0002-0073-0172}
\affiliation{
  \institution{City University of Hong Kong}
  \city{Hong Kong}
  \country{Hong Kong SAR}
}
\email{morin.wang@my.cityu.edu.hk}

\author{Yichao Wang}
\orcid{0000-0001-7053-8269}
\affiliation{
  \institution{Huawei Noah’s Ark Lab}
  \city{Shenzhen}
  \country{China}
}
\email{wangyichao5@huawei.com}

\author{Zhaocheng Du}
\orcid{0000-0002-1811-129X}
\affiliation{
  \institution{Huawei Noah’s Ark Lab}
  \city{Shenzhen}
  \country{China}
}
\email{zhaochengdu@huawei.com}

\author{Huifeng Guo}
\authornotemark[2]
\orcid{0000-0002-7393-8994}
\affiliation{
  \institution{Huawei Noah’s Ark Lab}
  \city{Shenzhen}
  \country{China}
}
\email{huifeng.guo@huawei.com}

\author{Yong Liu}
\orcid{0000-0001-9031-9696}
\affiliation{
  \institution{Huawei Noah’s Ark Lab}
  \city{Shenzhen}
  \country{China}
}
\email{liu.yong6@huawei.com}

\author{Ruiming Tang}
\orcid{0000-0002-9224-2431}
\affiliation{
  \institution{Huawei Noah’s Ark Lab}
  \city{Shenzhen}
  \country{China}
}
\email{tangruiming@huawei.com}

\author{Xiangyu Zhao}
\orcid{0000-0003-2926-4416}
\authornotemark[2]
\affiliation{
  \institution{City University of Hong Kong}
  \city{Hong Kong}
  \country{Hong Kong SAR}
}
\email{xianzhao@cityu.edu.hk}

\renewcommand{\shortauthors}{Yingyi Zhang et al.}

\newcommand{\zxy}[1]{{\color{blue} [#1 – Xy]}}
\newcommand{\zc}[0]{{\color{blue} [cite]}\xspace}
\newcommand{\yc}[1]{{\color{purple} [yichao: #1]}}
\newcommand{\py}[1]{{\color{red} [Pengyue: #1]}}
\newcommand{\xdr}[1]{{\color{orange} [xdr: #1]}}
\newcommand{\yyy}[1]{{\color{black} #1}}
\newcommand{\yy}[1]{{\color{black} #1}}
\begin{abstract}

Cloud-device collaboration leverages on-cloud Large Language Models (LLMs) for handling public user queries and on-device Small Language Models (SLMs) for processing private user data, collectively forming a powerful and privacy-preserving solution.
However, existing approaches often fail to fully leverage the scalable problem-solving capabilities of on-cloud LLMs while underutilizing the advantage of on-device SLMs in accessing and processing personalized data.
This leads to two interconnected issues: 
1) Limited utilization of the problem-solving capabilities of on-cloud LLMs, which fail to align with personalized user-task needs, 
and 2) Inadequate integration of user data into on-device SLM responses, resulting in mismatches in contextual user information.

In this paper, we propose a \textbf{L}eader–\textbf{S}ubordinate \textbf{R}etrieval framework for \textbf{P}rivacy-preserving cloud–device collaboration (LSRP), a novel solution that bridges these gaps by:
1) enhancing on-cloud LLM guidance to on-device SLM through a dynamic selection of task-specific leader strategies named as user-to-user retrieval-augmented generation ({U-U-RAG}),
and 2) integrating the data advantages of on-device SLMs through small model feedback Direct Preference Optimization ({SMFB-DPO}) for aligning the on-cloud LLM with the on-device SLM.
Experiments on two datasets demonstrate that {LSRP} consistently outperforms state-of-the-art baselines, significantly improving question-answer relevance and personalization, while preserving user privacy through efficient on-device retrieval. Our code is available at: \url{https://github.com/Applied-Machine-Learning-Lab/LSRP}.
\end{abstract}

\ccsdesc[500]{Computing methodologies~Natural language processing}
\ccsdesc[500]{Security and privacy~Privacy protections}
\ccsdesc[500]{Human-centered computing~Personal digital assistants}
\keywords{Cloud-Device Framework; Large Language Model; Privacy-Preserving}

\maketitle

\section{Introduction}

\begin{figure}
    \centering
    \includegraphics[width=1.0\linewidth]{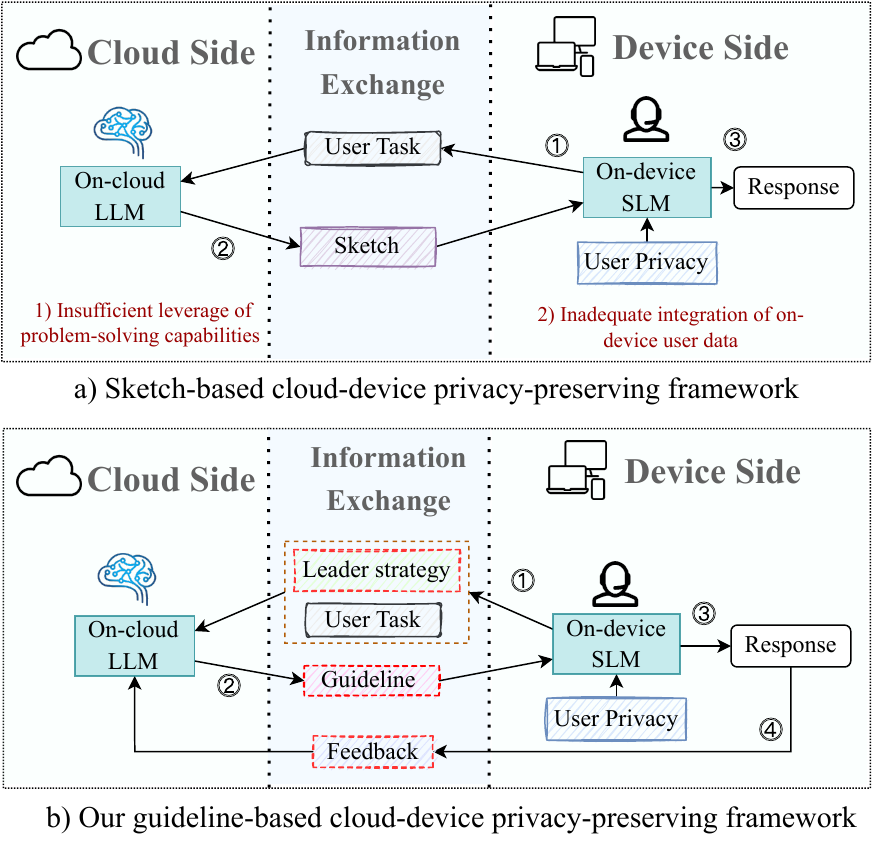}
    \caption{Illustration of the limitations of the device-cloud framework in privacy-preserving tasks.}
    \label{fig:intro_fig}
\end{figure}

Recently, enterprises leveraging Large Language Models (LLMs) have begun collecting extensive user data through question-and-answer interactions to enhance and personalize their services~\cite{chen2024large,wu2024understanding,liu2024large,gao2024llm,xu2024large,jia2023mill}.
However, the global regulatory landscape has become increasingly stringent in safeguarding user privacy, as exemplified by the European Union’s General Data Protection Regulation (GDPR)~\cite{regulation2016regulation}, which imposes strict requirements on the collection and use of personal data. 
In compliance with such regulations, private user data cannot be accessed outside the device, posing significant challenges for AI service providers.
Additionally, the computational constraints of devices further limit the feasibility of deploying LLMs on-device, complicating the development of privacy-preserving solutions.

In response~\cite{lu2024small}, major technology companies such as Apple, Google, and Huawei have introduced smaller, on-device Small Language Models (SLMs) to process sensitive user data locally. 
Notable examples include OpenELM-1.1B~\cite{mehtaOpenELMEfficientLanguage2024}, Gemma-2-2B~\cite{team2024gemma}, and PanGu-$\pi$-1B~\cite{tang2024rethinking}. However, with limited parameters, on-device SLMs often struggle with limited problem-solving capabilities compared to on-cloud LLMs, leading to a suboptimal user experience.
Therefore, technology companies have developed cloud-device collaboration frameworks that should integrate the problem-solving capabilities and privacy-preserving capabilities of on-cloud LLMs and on-device SLMs.
Current cloud-device collaborative frameworks primarily rely on cloud LLMs to generate a preliminary response sketch~\cite{cogenesis}. 
Specifically, on-device models send publicly shareable, non-private queries to the cloud, where the cloud model generates a sketch response. 
This response is then returned to the on-device SLM, which refines it to enhance the final output while preserving user privacy.

However, relying solely on the cloud model to provide sketch response fails to fully exploit the problem-solving advantages of LLMs and the data-specific strengths of SLMs. Specifically:
\textbf{1) Limited utilization of on-cloud LLMs' problem-solving capabilities}: On-cloud LLMs generate responses based on standardized prompts, which often fail to align with users' specific and personalized needs~\cite{li2025survey}. As a result, even after incorporating knowledge from the cloud model, the on-device SLM may still produce responses that lack sufficient relevance and personalization. This limitation hinders its ability to effectively handle diverse, complex, and highly personalized user tasks.
\textbf{2) Inadequate integration of on-device user data}: While cloud LLMs are constrained by privacy restrictions in processing user data due to privacy constraints, on-device SLMs have access to richer, privacy-sensitive data. However, cloud-generated responses often fail to incorporate the full user context available locally, limiting the SLM’s ability to capitalize on this data advantage. This misalignment further diminishes overall system performance.
Given these challenges, a key research question arises: \textit{How can cloud-device collaborative frameworks be designed to maximize the performance of on-device models while preserving user privacy and ensuring alignment with individual user needs?}

To fully harness the capabilities of both on-cloud LLMs and on-device SLMs, we propose a \textbf{L}eader–\textbf{S}ubordinate \textbf{R}etrieval framework for \textbf{P}rivacy-preserving cloud–device collaboration (LSRP).  Our goal is to enhance the performance of on-device SLM in a privacy-preserving setting by leveraging the problem-solving capabilities of on-cloud LLMs without exposing private user data. 
Figure~\ref{fig:intro_fig} illustrates our framework, which addresses the limitations of the sketch-based method, such as insufficient utilization of the LLM’s problem-solving capabilities and inadequate integration of SLM user data. 
Our method positions the on-cloud LLM as the leader to generate guidelines and fully leverage its capabilities, while the on-device SLM functions as the subordinate, following the path-goal theory~\cite{house1971path}. By working in tandem, these models collaborate to efficiently execute user tasks while balancing problem-solving capacity and privacy considerations. Specifically:
\textbf{1) Leveraging the problem-solving capability of on-cloud LLMs,} the framework employs multiple leadership roles to generate high-quality, task-specific guidelines, enhancing the personalization of on-device responses by guiding them with tailored input from the cloud following the path-goal theory~\cite{house1971path}.
Furthermore, recognizing that unified guidelines often fail to capture user preferences and adapt to diverse task requirements, we introduce a structured approach to cloud-generated guidance. Specifically, we conceptualize the guidance provided by the cloud LLM as a form of ``leadership behavior'' and propose a leadership-style-based guideline generation method. Under this framework, the system dynamically selects the most appropriate leader strategy for a specific user-task through our proposed user-to-user retrieval-augmented generation (U-U-RAG) method, thereby optimizing the effectiveness of the guidance. 

\textbf{2) Integrating the data advantages of on-device SLMs,} we incorporate feedback based on user-specific privacy-preserving metrics of question-answer relevance (Q-A Rel.), user data reference rate (UDRR), and perplexity (PPL) to optimize the on-cloud LLM, enhancing both response quality and personalization.
Existing cloud-generated drafts often fail to fully align with the inputs of on-device SLMs, limiting their effectiveness and personalization.
To address this issue, we introduce a small model feedback mechanism, wherein the on-cloud LLM generates two distinct guidelines for the same query. These guidelines are then executed and evaluated by the on-device SLM, which labels them as either “prefer” or “reject.” This feedback is subsequently used to fine-tune the on-cloud LLM via Direct Preference Optimization (DPO)~\cite{dpo}, improving its problem-solving capabilities and alignment with on-device needs. This iterative refinement process ensures that the cloud LLM effectively complements the data-rich context of on-device SLMs, significantly enhancing the adaptability and effectiveness of the collaborative framework.
Our contributions can be summarized as follows:
\begin{itemize}[leftmargin=*]
    \item We propose a leadership-style-based U-U-RAG generation method for on-cloud LLMs, leveraging RAG with user interactions to dynamically select the most suitable leadership strategies for specific tasks.
    \item We introduce a feedback-driven optimization method for on-cloud LLMs, utilizing feedback from on-device SLMs to enhance personalization and effectiveness, addressing a critical gap in optimizing cloud models within cloud-device frameworks. 
    \item Extensive experiments on two datasets validate the effectiveness of LSRP, demonstrating significant improvements in both relevance and personalization compared to baseline methods.
\end{itemize}

\section{Problem Definition}
Input-output-level privacy-preserving cloud-device LLM-SLM collaboration aims to leverage the output of on-cloud LLMs to enhance the performance of on-device SLMs. Let the user task be represented by $\mathcal{T}$, the user privacy by $\mathcal{P}$, the on-cloud LLM with parameters $\theta$, and the on-device SLM with parameters $\phi$.

In this collaboration, the on-cloud LLM processes the user task $\mathcal{T}$ and outputs an intermediate result \yyy{(\textit{e.g.}, query expand and write sketch)} denoted as $\mathcal{O}_{\mathrm{LLM}} = \mathcal{F}_{\mathrm{LLM}}(\mathcal{T}, \theta)$. This output serves as an input to the on-device SLM. The on-device SLM further processes the intermediate result, with user privacy $\mathcal{P}$ and its own parameters $\phi$ to generate the final output:
\begin{equation}
    \mathcal{O}_{\mathrm{SLM}} = \mathcal{F}_{\mathrm{SLM}}(\mathcal{O}_{\mathrm{LLM}}, \mathcal{P}, \phi).
\end{equation}

The primary objectives of this collaboration are:
\begin{enumerate}[leftmargin=*]
    \item \textbf{Privacy Preservation:} Ensuring user privacy $\mathcal{P}$ remains strictly on the device and is not transmitted to the cloud at any stage.
    \item \textbf{Performance Enhancement:} Utilizing the intermediate result $\mathcal{O}_{\mathrm{LLM}}$ from the on-cloud LLM to improve the quality and personalization of the on-device SLM output.
\end{enumerate}

The optimization goal of this collaboration system is to maximize user-task-specific performance $\mathcal{R}$:
\begin{equation}
\begin{aligned}
    \mathcal{R} = \max_{\theta} \mathcal{Q}\big(\mathcal{F}_{\mathrm{SLM}}(\mathcal{F}_{\mathrm{LLM}}(\mathcal{T}, \theta), \mathcal{P}, \phi)\big), \\
    s.t. \mathcal{P} \not\in \{\text{data transmitted to the cloud}\}.
\end{aligned}
\end{equation}
Here, $\mathcal{Q}(\cdot)$ represents the performance evaluation metric for the task $\mathcal{T}$, which will be detailed in the Section~\ref{sec:Q}.

\begin{figure*}[th]
    \centering
    \includegraphics[width=1.0\linewidth]{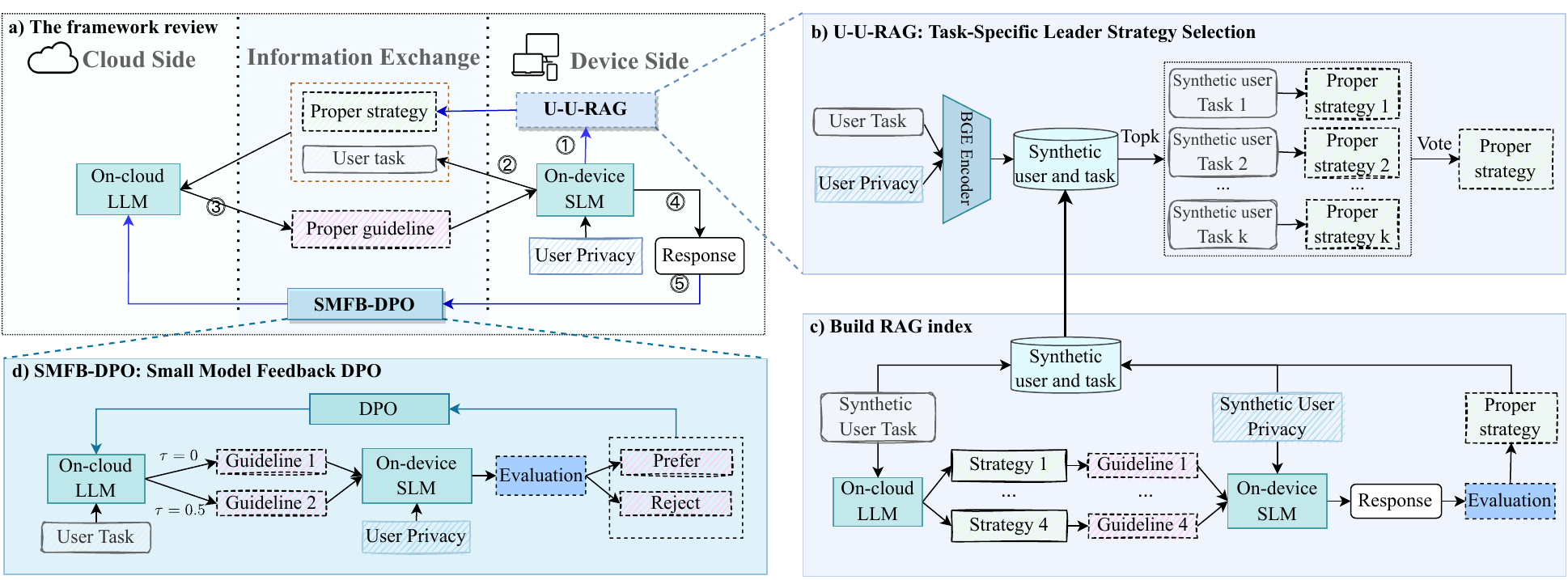}
    \caption{Main framework of LSRP. }
    \label{fig:main_framework}
\end{figure*}

\section{LSRP Framework}
\subsection{Leader-Subordinate Framework Overview}
To improve the effectiveness of cloud-device LLM-SLM collaboration, we propose a guideline-based leader-subordinate framework inspired by the \textit{path-goal leadership theory}~\cite{house1971path}. This theory is particularly well-suited for cloud-device collaboration as it reflects the relationship between a powerful leader, which provides guidance, and a less capable subordinate, which focuses on optimizing task execution. In this framework, the cloud LLM serves as the \textbf{leader}, offering guideline, while the on-device SLM functions as the \textbf{subordinate}, delivering privacy-preserving and personalized task execution as shown in Figure~\ref{fig:main_framework} a). The collaboration contains two-side:
\begin{itemize}[leftmargin=*]
    \item \textbf{On-Cloud LLM as Leader:} 
    The cloud LLM, equipped with strong problem-solving capability, generates high-quality intermediate outputs or task-specific guidance based on the user task $\mathcal{T}$. This guidance, denoted as $\mathcal{O}_{\mathrm{LLM}}$, serves as a structured pathway to guide the on-device SLM.
    \item \textbf{On-Device SLM as Subordinate:} The on-device SLM receives the intermediate output $\mathcal{O}_{\mathrm{LLM}}$ and combines it with user privacy $\mathcal{P}$ and its localized parameters $\phi$ to generate the final task output $\mathcal{O}_{\mathrm{SLM}}$. The SLM's role is to adapt the leader's guidance to the personalized and privacy-preserving requirements of the user.
\end{itemize}

The collaboration between the two-sided LLM and SLM follows five steps: Step \ding{192} On-device SLM selects the task-related leader strategy and transmits it to the cloud. Step \ding{193} On-device SLM transmits the user task to the cloud. Step \ding{194} The on-cloud LLM generates guidelines and transmits them to the on-device. Step \ding{195} The on-device SLM generates the final response based on the user task, user privacy, and guidelines. Step \ding{196} On-device evaluation is transmitted to optimize the LLM.
The primary problem can be expressed as an optimization of the cloud LLM's output $\mathcal{O}_{\mathrm{LLM}}$ to guide the on-device SLM effectively:
\begin{equation}
    \mathcal{R} = \max_{\mathcal{O}_{\mathrm{LLM}}} \mathcal{Q}\big(\mathcal{F}_{\mathrm{SLM}}(\mathcal{O}_{\mathrm{LLM}}, \mathcal{P}, \phi)\big),\ \text{where}\ \mathcal{O}_{\mathrm{LLM}} = \mathcal{F}_{\mathrm{LLM}}(\mathcal{T}, \theta).
\end{equation}

Thus, the focus is on optimizing the intermediate outputs $\mathcal{O}_{\mathrm{LLM}}$, \textit{i.e.}, the guidelines, to provide the most effective guidance for the on-device SLM. This is achieved by generating task-specific guidelines through a \textbf{task-specific leader strategy} and refining the parameters of the on-cloud LLM based on \textbf{small model feedback using DPO}, which will be discussed in the Section~\ref{U-U-RAG} and \ref{SMFB_DPO}.

\subsection{U-U-RAG: Task-Specific Leader Strategy Selection}
\label{U-U-RAG}
\yyy{The major challenge in cloud-device collaboration is handling the diverse and dynamic nature of tasks and user needs. Existing methods treat tasks uniformly and fail to adapt to task complexity or user preferences, resulting in insufficient leverage of the problem-solving capabilities of on-cloud LLMs.}

To address these issues, we propose that the on-cloud LLM adopt different leadership strategies depending on the task at hand. Drawing from \textit{path-goal leadership} theory, the cloud-based LLM, acting as the leader, employs four distinct leadership strategies to guide the on-device SLM (the subordinate). These strategies are designed to address the diverse needs of each task, ensuring that the guidance provided is tailored to both the task’s complexity and the user’s specific requirements. By leveraging the problem-solving advantages of the cloud model, this approach enables more personalized and context-aware responses, improving the collaboration between the on-cloud LLM and on-device SLM.

We define these four strategies to optimize task performance and personalization in a manner that is tailored to each unique scenario, with each strategy as a specific \textit{prompt} $\mathbb{P}_{\mathrm{LLM}}$.
The primary objective is to identify the most suitable leadership strategy for a given task $\mathcal{T}$ and user privacy $\mathcal{P}$. 
We designed the following four strategies to address the varying levels of task complexity and the degree of involvement required from the on-device models. The detailed prompts of four strategies can be found at our code \footnote{\url{https://github.com/Applied-Machine-Learning-Lab/LSRP}} and the Appendix~\ref{app:cogen_prompt} and \ref{app:movie_prompt}.
These strategies are defined as follows:
\begin{enumerate}[leftmargin=*]
    \item \textbf{Directive strategy:} The cloud LLM responds to a directive prompt $\mathbb{P}_{\mathrm{LLM}}^{\mathrm{dir}}$. This strategy is employed when the task requires clear, step-by-step guidance. The on-cloud LLM provides structured instructions for the on-device SLM, ensuring the task is completed with precision.
    \item \textbf{Supportive strategy:} The cloud LLM processes a supportive prompt $\mathbb{P}_{\mathrm{LLM}}^{\mathrm{sup}}$. When the task allows for creativity and flexibility, this strategy is used to encourage the on-device SLM to explore multiple solutions. The cloud LLM offers general direction and moral support, allowing the on-device model to personalize the task further.
    \item \textbf{Participative strategy:} The on-device SLM interacts with a participative prompt $\mathbb{P}_{\mathrm{LLM}}^{\mathrm{par}}$. The cloud LLM provides guidelines while allowing the on-device model to make significant contributions, ensuring the final output is highly personalized. 
    \item \textbf{Achievement-Oriented strategy:} The cloud LLM processes an achievement-oriented prompt $\mathbb{P}_{\mathrm{LLM}}^{\mathrm{ach}}$. This strategy is applied to tasks that require high performance and excellence. The cloud LLM sets high standards and challenges the on-device model to exceed expectations.
\end{enumerate}

These strategies were designed to address tasks with different levels of complexity and varying degrees of involvement from the on-device model. The leadership approach is chosen based on the task's needs, ensuring that the cloud LLM provides appropriate guidance to the on-device SLM, thus enhancing overall collaboration and task performance.

To select the most proper leader strategy for real users, motivated by the principle ``Birds of a feather flock together''~\cite{ip2006birds}, which is largely used in recommender system field~\cite{crandall2008feedback, zhao2018deep,zhao2018recommendations,fu2023unified, wang2023plate},  we propose a user-to-user retrieval argument generation (U-U-RAG) method to retrieve the most suitable guideline for a given user-task, as shown in  Figure~\ref{fig:main_framework} b). 

\subsubsection{\textbf{Index Building}}
As shown in Figure~\ref{fig:main_framework} c), we first build the RAG index by using synthetic users.
First, we construct synthetic users, each characterized by their privacy data $\mathcal{P}_{\text{syn}}$ and task $\mathcal{T}_{\text{syn}}$. For each synthetic user-task pair $(\mathcal{P}_{\text{syn}}, \mathcal{T}_{\text{syn}})$, the cloud LLM generates four distinct guidelines $\{\mathcal{O}_{\mathrm{LLM}}^{\text{Lead}}\}$, corresponding to the four leader strategies:
\begin{equation}
    \mathcal{O}_{\mathrm{LLM}}^{\text{Lead}} = \mathcal{F}_{\mathrm{LLM}}(\mathcal{T}_{\text{syn}}, \mathbb{P}_{\mathrm{LLM}}^{\text{Lead}}, \theta), \text{Lead} \in \{dir, sup, par, ach\},
\end{equation}
where $\mathbb{P}_{\mathrm{LLM}}^{\text{Lead}}$ represents the prompt for a specific leader strategy.
Next, the on-device SLM processes each guideline $\mathcal{O}_{\mathrm{LLM}}^{\text{Lead}}$, along with the synthetic user's privacy data $\mathcal{P}_{\text{syn}}$ and its parameters $\phi$, to generate the final output, with each evaluated using a metric $\mathcal{Q}(\cdot)$. And the most suitable guideline is determined by selecting the proper strategy that maximizes $\mathcal{R}^{\text{Lead}}$.
\begin{equation}
    \begin{aligned}
        \mathcal{O}_{\mathrm{SLM}}^{\text{Lead}} &= \mathcal{F}_{\mathrm{SLM}}(\mathcal{O}_{\mathrm{LLM}}^{\text{Lead}}, \mathcal{P}_{\text{syn}}, \phi),\\
        \mathcal{R}^{\text{Lead}} &= \mathcal{Q}\big(\mathcal{O}_{\mathrm{SLM}}^{\text{Lead}}\big), \quad \text{Lead} \in \{dir, sup, par, ach\}, \\
        \mathbb{P}_{\mathrm{LLM}}^{\text{Lead}^*} &= \arg\max_{\text{Lead}} \mathcal{R}^{\text{Lead}}.
    \end{aligned}
\end{equation}

Subsequently, we use a text encoder to vectorize the user-task pair $(\mathcal{P}_{\text{syn}}, \mathcal{T}_{\text{syn}})$.
\begin{equation}
    \mathbf{e}_{\text{syn}} = \mathrm{Encode}([\mathcal{P}_{\text{syn}}, \mathcal{T}_{\text{syn}}]),
\end{equation}
where $\mathrm{Encode}(\cdot)$ denotes the embedding function. These embeddings and the selected strategy $\mathbb{P}_{\mathrm{LLM}}^{\text{Lead}^*}$ are then stored in an embedding database $\mathcal{D}_{\text{embed}}$ for future retrieval.

\subsubsection{\textbf{Strategy Retrieval}} 
For a real user-task pair $(\mathcal{P}, \mathcal{T})$, we retrieve the most relevant synthetic user-task pair using RAG $(\mathcal{P}_{\text{syn}}, \mathcal{T}_{\text{syn}})$ . The following steps are performed: 

\textbf{(1) Encoding}. The real user-task pair is vectorized using the same encoder:
\begin{equation}
    \mathbf{e}_{\text{real}} = \mathrm{Encode}(\mathcal{P}, \mathcal{T}).
\end{equation}

\yyy{\textbf{(2) Retrieval.}} The top-$k$ most similar synthetic user-task pairs and their corresponding $\mathbb{P}_{\mathrm{LLM}}^{\text{Lead}^*}$ are retrieved from the embedding database $\mathcal{D}_{\text{embed}}$ using a similarity measure $\mathrm{Sim}(\cdot, \cdot)$:
\begin{equation}
    \mathcal{R}_{\text{top-}k} = \arg\max_{\mathbf{e} \in \mathcal{D}_{\text{embed}}} \mathrm{Sim}(\mathbf{e}_{\text{real}}, \mathbf{e}),
\end{equation}
where $\mathcal{R}_{\text{top-}k}$ is the set of the top-$k$ related user-task embeddings.

\yyy{\textbf{(3) Voting.}} The strategies associated with the retrieved $\mathbb{P}_{\mathrm{LLM}}^{\text{Lead}^*}$ are used for voting to determine the most suitable strategy for the real user-task pair. The final selected strategy $\mathbb{P}_{\mathrm{LLM}}^{(*)}$ is given by:
\begin{equation}
    \mathbb{P}_{\mathrm{LLM}}^{\text{Lead}^*} = \mathrm{Vote}(\{\mathbb{P}_{\mathrm{LLM}}^{\text{Lead}} \mid \mathbf{e}^{\text{Lead}} \in \mathcal{R}_{\text{top-}k}\}),
\end{equation}
where $\mathrm{Vote}(\cdot)$ aggregates the strategies based on their occurrences among the top-$k$.
This approach ensures that the most appropriate leader strategy is selected by leveraging user-to-user retrieval and a voting mechanism.

\subsection{SMFB-DPO: Small Model Feedback DPO}
\label{SMFB_DPO}
\yyy{Existing methods often overlook the importance of feedback from on-device SLMs which have data advantage. These feedback signals, which reflect user-specific context and task intricacies, are frequently ignored, resulting in suboptimal performance in real-world applications.} 
Inspired by \textit{path-goal leadership theory}~\cite{house1971path}, which suggests that a leader improves performance by incorporating feedback from subordinates, we propose the Small Model Feedback DPO (SMFB-DPO) method to optimize the parameters of the on-cloud LLM.

The process begins with the on-cloud LLM generating two guidelines $\{\mathcal{O}_{\mathrm{LLM}}^{j}\}_{j \in \{0,1\}}$ with different temperature $\tau$ for a given user task $\mathcal{T}$, as shown in Figure~\ref{fig:main_framework} d). 
In each feedback cycle, the number of samples fed back by these four strategies varies, and this imbalance can introduce bias into the model’s learning. To address this, we have designed a general strategy for DPO on-cloud LLMs to enhance the generalization of the DPO parameters.
\begin{equation}
    \mathbb{P}_{\mathrm{LLM}} = General(\mathbb{P}_{\mathrm{LLM}}^{\text{lead}}), \quad \text{lead} \in \{dir, sup, par, ach\},
\end{equation}
where $General(\mathbb{P}_{\mathrm{LLM}}^{\text{lead}})$ summarizes the general leader strategy prompt derived from the four leader strategies.

Each generated guideline $\mathcal{O}_{\mathrm{LLM}}^{j}$ is then input to the on-device SLM. The on-device SLM processes the guideline and produces corresponding outputs. These outputs are assessed based on the metric $\mathcal{Q}(\cdot)$, which results in feedback scores for each guideline. Based on the feedback scores $\mathcal{R}^{j}$, the on-device SLM identifies the most effective guideline by selecting the one with the highest score.
\begin{equation}
    \begin{aligned}
        \mathcal{R}^{j} &= \mathcal{Q}\big(\mathcal{O}_{\mathrm{SLM}}^{j}\big), \ \text{where} \ \mathcal{O}_{\mathrm{SLM}}^{j} = \mathcal{F}_{\mathrm{SLM}}(\mathcal{O}_{\mathrm{LLM}}^{j}, \mathcal{P}, \phi),\\
        p &= \arg\max_{j \in \{0,1\}} \mathcal{R}^{j}.
    \end{aligned}
\end{equation}

The selected guideline $\mathcal{O}_{\mathrm{LLM}}^{p}$ is labeled as ``prefer'', while the remaining one is labeled as ``rejec'' $\mathcal{O}_{\mathrm{LLM}}^{r}$.

The on-cloud LLM parameters $\theta$ are then updated using DPO to maximize the likelihood of generating the preferred guideline while minimizing the likelihood of generating rejected guidelines. The optimization objective is defined as:
\begin{equation}
    \mathcal{L}_{\mathrm{DPO}}(\theta) = - \sum_{N} \log P(\mathcal{O}_{\mathrm{LLM}}^{p} \mid \mathcal{T}, \theta) + \lambda \sum_{N} \log P(\mathcal{O}_{\mathrm{LLM}}^{r} \mid \mathcal{T}, \theta),
\end{equation}
\begin{equation}
    \theta^* = \arg\min_{\theta} \mathcal{L}_{\mathrm{DPO}}(\theta),
\end{equation}
where $\lambda$ is a weighting parameter controlling the balance between the preferred and rejected guidelines.
Finally, the fine-tuned on-cloud LLM will generate updated guidelines based on the new parameters $\theta^*$, which are better aligned with user preferences.

\subsection{Matrix $\mathcal{Q}$}
\label{sec:Q}
Ensuring that user privacy data is effectively utilized while maintaining high relevance in responses is a complex challenge. Developing metrics that assess on-device SLM's privacy-preserving data utilization without compromising performance or relevance~\cite{jia2024bridging} is particularly difficult. To evaluate the performance of the proposed framework on-device, we employ three metrics: Q-A Relevance (Q-A Rel. ↑), User Data Reference Rate (UDRR ↑), and Perplexity (PPL ↓), which measure the relevance of question-answer pairs, the utilization of user privacy data, and the perplexity of the SLM, respectively. 


\textbf{Q-A Relevance (Q-A Rel.)} measures the relevance of the on-device SLM's answer to the user task. This metric is inspired by the \textit{LLM-as-judge} framework~\cite{zheng2023judging} and aligns with the current practice of using LLMs as evaluators in cloud-device frameworks~\cite{cogenesis}.  For a given user task $\mathcal{T}$ and the corresponding output $\mathcal{O}_{\mathrm{SLM}}$, the relevance score $\mathcal{R}_{\mathrm{Q-A}}$ is computed as:
\begin{equation}
    \mathcal{R}_{\mathrm{Q-A}} = \mathrm{Score}_{\mathrm{LLM}}(\mathcal{T}, \mathcal{O}_{\mathrm{SLM}}),
\end{equation}
where $\mathrm{Score}_{\mathrm{LLM}}$  is a function that assigns a relevance score based on semantic similarity and task alignment.

\textbf{User Data Reference Rate (UDRR)} measures the proportion of user-specific information from the user privacy data that is referenced in the output of SLM. This metric provides a quantitative assessment of how effectively the on-device model leverages user-specific information in its generated outputs, ensuring personalized responses while maintaining privacy-preserving principles. For a single instance, the UDRR is defined as:
\begin{equation}
    \mathrm{UDRR}_{i} = \frac{|\mathcal{W}_{\mathrm{user}} \cap \mathcal{W}_{\mathrm{output}}|}{|\mathcal{W}_{\mathrm{user}}|},
\end{equation}
where $\mathcal{W}_{\mathrm{user}}$ is the set of unique words in the user data $\mathcal{P}$, $\mathcal{W}_{\mathrm{output}}$ is the set of unique words in the on-device model's generated output $\mathcal{O}_{\mathrm{SLM}, i}$, and $|\cdot|$: The cardinality (number of elements) of a set.

\textbf{Perplexity (PPL)} evaluates the fluency and coherence of the on-device SLM's generated outputs. For a generated output $\mathcal{O}_{\mathrm{SLM}}$, the perplexity is defined as:
\begin{equation}
    \mathrm{PPL} = \exp\left(-\frac{1}{N} \sum_{i=1}^N \log P(w_i \mid w_1, w_2, \dots, w_{i-1})\right),
\end{equation}
where $P(w_i \mid w_1, w_2, \dots, w_{i-1})$ represents the conditional probability of token $w_i$ given its preceding tokens, and $N$ is the number of tokens in the output.

\textbf{Metric $\mathcal{Q}(\cdot)$.} We construct a comprehensive metric $\mathcal{Q}(\cdot)$ as a weighted combination of the three evaluation metrics. The overall evaluation score for a given SLM response $\mathcal{O}_{\mathrm{SLM}}$ is defined as:
\begin{equation}
    \mathcal{Q}(\mathcal{O}_{\mathrm{SLM}}) = w_1 \cdot \mathcal{R}_{\mathrm{Q-A}} + w_2 \cdot \mathrm{UDRR} - w_3 \cdot \mathrm{PPL}.
\end{equation}

To determine the optimal weights $(w_1, w_2, w_3)$, we employ the NSGA-II algorithm~\cite{deb2002fast}, which performs multi-objective optimization to balance the trade-offs between Q-A Rel., UDRR, and PPL. Specifically, we define the objective vector as:
\begin{equation}
    \mathbf{f} = \big(-\mathcal{R}_{\mathrm{Q-A}}, -\mathrm{UDRR}, \mathrm{PPL}\big),
\end{equation}
where the negative signs for $\mathcal{R}_{\mathrm{Q-A}}$ and $\mathrm{UDRR}$ ensure that higher values of these metrics are treated as desirable objectives to maximize. The NSGA-II algorithm then identifies the Pareto front $\mathcal{F}$, consisting of non-dominated solutions in the objective space. Each point in the Pareto front is a $(w_1, w_2, w_3)$. From the Pareto front, we select the \textbf{knee point}~\cite{zhang2014knee}, which represents the optimal trade-off among the three metrics. 

\section{Experiment}
In this section, we conduct experiments to address the following research questions:
\begin{itemize}[leftmargin=*]
    \item \textbf{RQ1:} Can the leader-subordinate collaboration framework enhance the relevance and personalization of the on-device model's responses?
    \item \textbf{RQ2:} What is the specific impact of U-U-RAG and SMFB-DPO on the performance improvement of the on-device model?
    \item \textbf{RQ3:} Does the proposed U-U-RAG leader strategy leverage of problem-solving capabilities of on-cloud LLMs?
    \item \textbf{RQ4:} Does the proposed SMFB-DPO enhance the utilization of on-device user data and improve personalization?
\end{itemize}

\subsection{Experiment Settings}
\subsubsection{\textbf{Dataset}}
Due to the limitations of open-source data in open privacy, we use the following datasets: one is an open-source CoGensis dataset, and the other is a synthetic Movie Explain dataset created for this study, with statistic information shown in Table~\ref{tab:dataset_description}.
\begin{itemize}[leftmargin=*]
    \item \textbf{CoGensis dataset.} The recently proposed CoGensis~\cite{cogenesis} dataset includes user privacy data and a diverse range of personalized tasks. Each task involves long-text writing, requiring the integration of user-specific information during the process. The training-to-testing split is followed by this work~\cite{cogenesis}.
    \item \textbf{Movie Explain dataset.} Using synthetic user data~\cite{yukhymenko2024synthetic}, we create a movie recommendation explanation dataset. Specifically, we employ advanced LLM to generate movie recommendations for each user based on their private personal data, detailed in the Appendix~\ref{sec:data_gen}. Subsequently, we use the cloud-device framework to generate personalized explanations tailored to the user's data, with a training-to-testing ratio of 8:2.
\end{itemize}

\begin{table}[ht]
    \centering
    \caption{\yyy{Dataset description}}
    \vspace{-10pt}
    \resizebox{80mm}{!}{
    \begin{tabular}{lccc}
        \toprule
        \textbf{Dataset}& \textbf{Assumption} & \textbf{CoGenisis} & \textbf{Movie Explain} \\ \midrule
        \# training set & synthetic user-task & 1,346 & 959 \\ \hline
        \# test set & real user-task& 240 & 240 \\ \bottomrule
    \end{tabular}
    }
    \vspace{-10pt}
    \label{tab:dataset_description}
\end{table}

\begin{table*}[t]
    \centering
    \caption{This table presents the performance comparison of our method, 	LSRP, against baseline approaches on the CoGen Dataset (diverse tasks) and the Movie Explain Dataset (single task). The standard variance within the test set indicates the consistency of performance, demonstrating the robustness of our method. ``\textbf{{\Large *}}'' indicates the statistically significant improvements (i.e., two-sided t-test with $p<0.05$) over the best cloud-device baseline. $\uparrow$: higher is better.}
    \renewcommand{\arraystretch}{1} 
    \resizebox{165mm}{!}{
    \begin{tabular}{llcccc}
        \toprule
        & & \multicolumn{2}{c}{\textbf{LLaMa Series}}& \multicolumn{2}{c}{\textbf{Qwen Series}}\\ \cmidrule(rl){3-4} \cmidrule(rl){5-6}
        \textbf{Dataset}&\textbf{Framework}& \textbf{Q-A Rel.↑} & \textbf{Persona.↑ }&  \textbf{Q-A Rel.↑} & \textbf{Persona.↑} \\ \midrule
        \multirow{6}{*}{\textbf{CoGen Dataset}} &On-device SLM& 8.9383$\pm$4.0649 & 7.5166$\pm$1.5889  & 9.0750$\pm$1.5365 & 6.9041$\pm$1.8266 \\ 
        &Distillation & 8.5541$\pm$6.3387 & 7.5291$\pm$1.6525  & 8.9765$\pm$3.2605 & 6.7625$\pm$2.4422 \\ 
        &CoGenesis  & 6.1583$\pm$2.8082 & 5.8250$\pm$2.4821   & 6.6541$\pm$3.1598 & 5.4041$\pm$2.3697 \\ 
        &LLM guide SLM (ours) & 9.0125$\pm$2.4790 & 7.4500$\pm$2.6391  & 9.1483$\pm$1.2516 & 7.1580$\pm$1.6857 \\ 
        &\textbf{LSRP} & \textbf{9.2458$\pm$1.2950{\Large *}} & \textbf{7.5458$\pm$1.5213{\Large *}} & \textbf{9.3083$\pm$0.9729{\Large *}} & \textbf{7.2041$\pm$1.6309{\Large *}} \\
        &On-device LLM$^\dagger$  & 9.9791$\pm$0.0203 & 8.8166$\pm$0.8008 & 9.9666$\pm$0.2013 & 8.6708$\pm$0.8241 \\ \midrule
        \multirow{6}{*}{\textbf{Movie Explain Dataset}}&On-device SLM  & 7.6333$\pm$6.8822 & 6.8750$\pm$2.2915 & 7.9291$\pm$6.4991 & 7.3250$\pm$2.1628 \\ 
        &Distillation & 8.2666$\pm$2.3738 & 7.4416$\pm$1.2130  & 8.1541$\pm$3.8745 & 7.4833$\pm$1.3290 \\ 
        &CoGenesis  & 6.2291$\pm$4.0099 & 5.6166$\pm$3.2124  & 7.1125$\pm$4.4081 & 6.4458$\pm$2.6166 \\ 
        &LLM guide SLM (ours) & 8.5500$\pm$3.7475 & 7.7500$\pm$1.8062  & 8.2066$\pm$3.4872 & 7.5066$\pm$1.9005 \\ 
        &\textbf{LSRP} & \textbf{8.6708$\pm$2.5124{\Large *}} & \textbf{7.9750$\pm$1.4431{\Large *}} & \textbf{8.4521$\pm$2.0495{\Large *}} & \textbf{7.5541$\pm$1.7880{\Large *}}\\
        &On-device LLM$^\dagger$ & 9.8958$\pm$0.1016 & 9.2416$\pm$0.4999 &  9.8208$\pm$0.1637 & 9.3458$\pm$0.5255 \\ \bottomrule
    \end{tabular}
    }
    \label{tab:main_results}
    \begin{tablenotes}
        \small
        \item \textbf{Note:} On-device LLM$^\dagger$ represents the theoretical upper bound for Q-A task performance by deploying the LLM directly on the device. However, this is often infeasible for resource-constrained devices due to the high computational and storage requirements.
    \end{tablenotes}
\end{table*}

\begin{table*}[!ht]
    \centering
    \caption{Ablation study of our LSRP framework.}
    \resizebox{165mm}{!}{
    \begin{tabular}{llcclcc}
    \toprule
        \multirow{2}{*}{\textbf{Dataset}} &  \multicolumn{3}{c}{\textbf{LLaMa Series}}& \multicolumn{3}{c}{\textbf{Qwen Series}}\\ \cmidrule(rl){2-4} \cmidrule(rl){5-7}
        ~ & \textbf{Framework} & \textbf{Q-A Rel.↑} & \textbf{Persona↑ }& \textbf{Framework }& \textbf{Q-A Rel.↑} & \textbf{Persona↑} \\ \midrule
        \multirow{4}{*}{\textbf{CoGen Dataset}} & \textbf{LSRP} & \textbf{9.2458$\pm$1.2950} & \textbf{7.5458$\pm$1.5213} & \textbf{LSRP} & \textbf{9.3083$\pm$0.9729} & \textbf{7.2041$\pm$1.6309}  \\ 
        ~ & -wo U-U-RAG & 9.0666$\pm$2.6788 & 7.5291$\pm$2.4074 & -wo U-U-RAG & 9.1958$\pm$1.3659 & 7.1783$\pm$1.6628  \\ 
        ~ & -wo SMFB-DPO & 9.1750$\pm$1.5607 & 7.4875$\pm$1.3695 & -wo SMFB-DPO & 9.2000$\pm$1.3577 & 7.1745$\pm$1.8575  \\
        ~ & LLM guide SLM & 9.0125$\pm$2.4790 & 7.4500$\pm$2.6391 & LLM guide SLM & 9.1483$\pm$1.2516 & 7.1580$\pm$1.6857  \\ \midrule
        \multirow{4}{*}{\textbf{Movie Explain Dataset}} & \textbf{LSRP} & \textbf{8.6708$\pm$2.5124 }& \textbf{7.9750$\pm$1.4431} & \textbf{LSRP} & \textbf{8.4521$\pm$2.0495} & \textbf{7.5541$\pm$1.7880}  \\ 
        ~ & -wo U-U-RAG & 8.5916$\pm$2.4915 & 7.7652$\pm$1.7323 & -wo U-U-RAG & 8.2125$\pm$3.5006 & 7.5166$\pm$1.8402  \\ 
        ~ & -wo SMFB-DPO & 8.6166$\pm$2.5363 & 7.8375$\pm$1.4122 & -wo SMFB-DPO & 8.4416$\pm$2.6132 & 7.5416$\pm$1.7511  \\ 
        ~ & LLM guide SLM & 8.5500$\pm$3.7475 & 7.7500$\pm$1.8062 & LLM guide SLM & 8.2066$\pm$3.4872 & 7.5066$\pm$1.9005 \\ \bottomrule
    \end{tabular}
    }
    \label{tab:ablation_results}
\end{table*}
\subsubsection{\textbf{Baselines}}%
To evaluate the superiority of our proposed method, we compare LSRP with several baseline methods, as summarized below: \textbf{On-Device SLM.} This baseline uses only the on-device SLM to answer questions directly without any assistance from a cloud-based LLM. While ensuring strong privacy protection, it relies solely on the limited capabilities of the SLM.
\textbf{Distillation~\cite{wang2024large}.} In this method, the on-device SLM is fine-tuned using answers distilled from a cloud-based LLM. Although this approach enhances the SLM's performance, the fine-tuning process retains user privacy data, potentially leading to privacy leakage.
\textbf{CoGenesis~\cite{cogenesis}.} This sketch-based cloud-device framework utilizes the cloud LLM to generate a task sketch and the on-device SLM to complete it. While this method balances the strengths of both models, its reliance on sketches limits the flexibility and performance of the framework.
\textbf{LLM Guide SLM.} This is the basic version of our proposed guideline-based cloud-device framework. It uses general leader prompts in the cloud LLM to guide the on-device SLM in completing tasks. Although effective, it lacks task-specific optimization strategies.
\textbf{On-Device LLM.} This baseline uses only the cloud-based LLM to answer questions directly. While this achieves high performance due to the LLM's capabilities, it requires high local computation resources.

\subsubsection{\textbf{Implementation Detail}}
We use 4 advanced GPUs. The LLaMa series used are LLaMa-3.3-70B-Instruct and LLaMa-3.2-1B-Instruct~\cite{dubey2024llama}
The Qwen series used are Qwen2.5-72B-Instruct and Qwen2.5-1.5B-Instruct~\cite{yang2024qwen2}. For all inferences, we set the temperature to 0. More details can be seen in the Appendix \ref{app:imp_det}, and the latency details can be found at Appendix \ref{app:latency}.

\subsubsection{\textbf{Evaluation Metrics}}
Following previous work~\cite{cogenesis}, we employ two LLM-as-a-judge~\cite{zheng2023judging} metrics: \textbf{Q-A Relevance (Q-A Rel.)}, \textbf{Q-A Personalization (Persona)}, with prompts that can be found in the Appendix~\ref{app:eval_prompt}. We evaluate readability using the Flesch Reading Ease (FRE) score~\cite{readability}, with the results presented in Appendix~\ref{app:fre}. Each metric measures a specific aspect of the system's performance, as described below:
\yyy{
\begin{itemize}[leftmargin=*]
    \item \textbf{Q-A Relevance (Q-A Rel.)} measures the relevance of the on-device SLM's answer to the user task. This metric is inspired by the \textit{LLM-as-judge} framework~\cite{zheng2023judging} and aligns with the current practice of using LLMs as evaluators in cloud-device frameworks~\cite{cogenesis}. We leverage the advanced LLM to quantify the relevance. 
    \item \textbf{Q-A Personalization (Persona)} measures the degree to which the on-device SLM's answer is tailored to the user's profile, preferences, and contextual needs. We also leverage the advanced LLM to quantify personalization.

\end{itemize}
}


\label{sec:full_slm_choose}
\begin{table*}[!th]
    \centering
    \caption{Our LSRP performance using different leader strategy selection.}
    \resizebox{165mm}{!}{
    \begin{tabular}{llcclcc}
    \hline
         \multirow{2}{*}{\textbf{Dataset}} &  \multicolumn{3}{c}{\textbf{LLaMa Series}}& \multicolumn{3}{c}{\textbf{Qwen Series}}\\ \cmidrule(rl){2-4} \cmidrule(rl){5-7}
        ~ & \textbf{Framework} & \textbf{Q-A Rel.↑} & \textbf{Persona.↑ }& \textbf{Framework }& \textbf{Q-A Rel.↑} & \textbf{Persona.↑} \\ \midrule
        \multirow{3}{*}{\textbf{CoGen Dataset}}& LLM guide SLM & 9.0125$\pm$2.4790 & 7.4500$\pm$2.6391 & LLM guide SLM & 9.1483$\pm$1.2516 & 7.1580$\pm$1.6857  \\ 
        ~ & SLM-choose & 9.1416$\pm$1.4130 & 7.4833$\pm$1.6200 & SLM-choose & 9.1500$\pm$2.2308 & 7.1316$\pm$1.7948  \\ 
        ~ & U-U-RAG & \textbf{9.1750$\pm$1.5607} & \textbf{7.4875$\pm$1.3695} & U-U-RAG & \textbf{9.2000$\pm$1.3577 }& \textbf{7.1745$\pm$1.8575}  \\ 
        \midrule
        \multirow{3}{*}{\textbf{Movie Explain Dataset}} & LLM guide SLM & 8.5500$\pm$3.7475 & 7.7500$\pm$1.8062 & LLM guide SLM & 8.2066$\pm$3.4872 & 7.5066$\pm$1.9005  \\ 
        ~ & SLM-choose & 8.3958$\pm$3.6724 & 7.7125$\pm$1.7573 & SLM-choose & 8.4416$\pm$2.6132 & 7.5041$\pm$1.7051  \\ 
        ~ & U-U-RAG & \textbf{8.6166$\pm$2.5363} & \textbf{7.8375$\pm$1.4122} & U-U-RAG & \textbf{8.4521$\pm$2.0495} & \textbf{7.5166$\pm$1.8402 }\\ 
        \bottomrule
    \end{tabular}
    }
    \label{tab:choose_results}
\end{table*}

\subsection{Overall Performance (RQ1)}
To verify the effectiveness of our method, we conduct experiments on the CoGen and Movie Explain datasets with other baselines using LLaMa Series and Qwen Series models, with results shown in Table~\ref{tab:main_results}. Based on the results shown, we draw the following conclusions: 

\textbf{1) Leader-subordinate collaboration is important.} The results confirm that the leader-subordinate collaboration framework in LSRP significantly enhances the responses of on-device SLMs by improving personalization and relevance across both diverse and single-task datasets. And even the basic version, ``LLM Guide SLM,'' outperforms the sketch-based CoGenisis method.  Using U-U-RAG and SMFB-DPO, our LSRP framework surpasses the ``LLM Guide SLM,'' due to its ability to search for an appropriate leader strategy to generate guidelines and learn from subordinate feedback. A lower variance indicates that our method enables the samples to achieve consistently high scores.
\textbf{2) Leader-subordinate collaboration further improves Q-A Rel. in diverse tasks.} Comparing LSRP frameworks on the CoGenisis and Movie Explain datasets, we observe the results in a higher Q-A Rel. score of 9.3083 on the CoGenisis dataset, compared to 8.6708 on the Movie Explain dataset. The Q-A Rel. performance on the CoGenisis dataset can be attributed to the diversity of tasks, which offers greater potential for identifying a more suitable strategy. 
\textbf{3) Leader-subordinate collaboration further improves Persona in single tasks.} In contrast, the Persona score is higher on the Movie Explain dataset (7.9750) compared to the CoGenisis dataset (7.5458), indicating that our framework provides more personalized responses for a single task. The improved Persona performance on the Movie Explain dataset can be attributed to task consistency, which offers greater potential for refining the guideline to better utilize user privacy data.
We also conduct experiments on different scales of on-cloud LLMs and on-device SLMs, these results can be seen in the Appendix~\ref{sec:Diff_onc_ond}. A case study on comparing our LSRP and LLM guide SLM is also given in Appendix~\ref{sec:case} for reference.

\subsection{Ablation Study (RQ2)}
To understand the specific impact of each component in our LSRP framework, we design the following variants: (1) -wo U-U-RAG, (2) -wo SMFB-DPO, and (3) LLM guide SLM indicate without both. From Table~\ref{tab:ablation_results}, we can draw the following conclusions: 
\begin{itemize}[leftmargin=*]
    \item \textbf{1) Both U-U-RAG and SMFB-DPO are important.} The full LSRP framework achieves the highest Q-A Rel. and Persona scores. These findings highlight the synergistic impact of both components in delivering superior results.
    \item \textbf{2) U-U-RAG contributes more to Q-A Rel.} For the LLaMa and Qwen series, removing U-U-RAG results in a slight drop in Q-A Rel. and Persona. 
    \item \textbf{3) SMFB-DPO contributes more to Persona.} Similarly, removing SMFB-DPO leads to a decrease in Persona scores while marginally affecting Q-A Rel..
\end{itemize}
\begin{figure}[ht]
    \centering
    \includegraphics[width=0.99\linewidth]{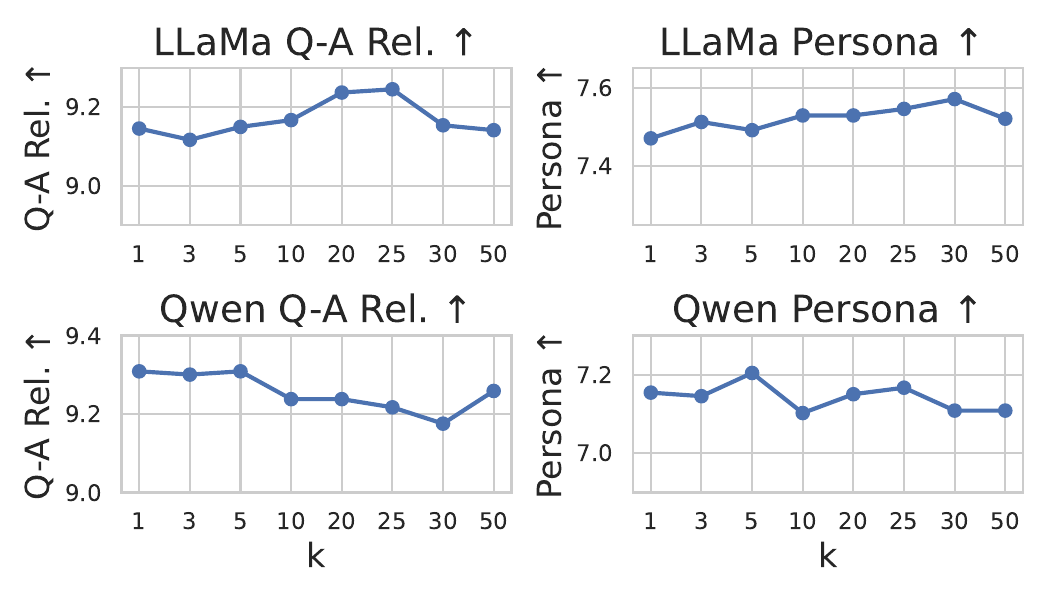}
    \vspace{-15pt}
    \caption{\yyy{Experiment on different $k$ settings.}}
    \vspace{-15pt}
    \label{fig:sensi_k}
\end{figure}
\begin{figure}[ht]
    \centering
    \includegraphics[width=0.99\linewidth]{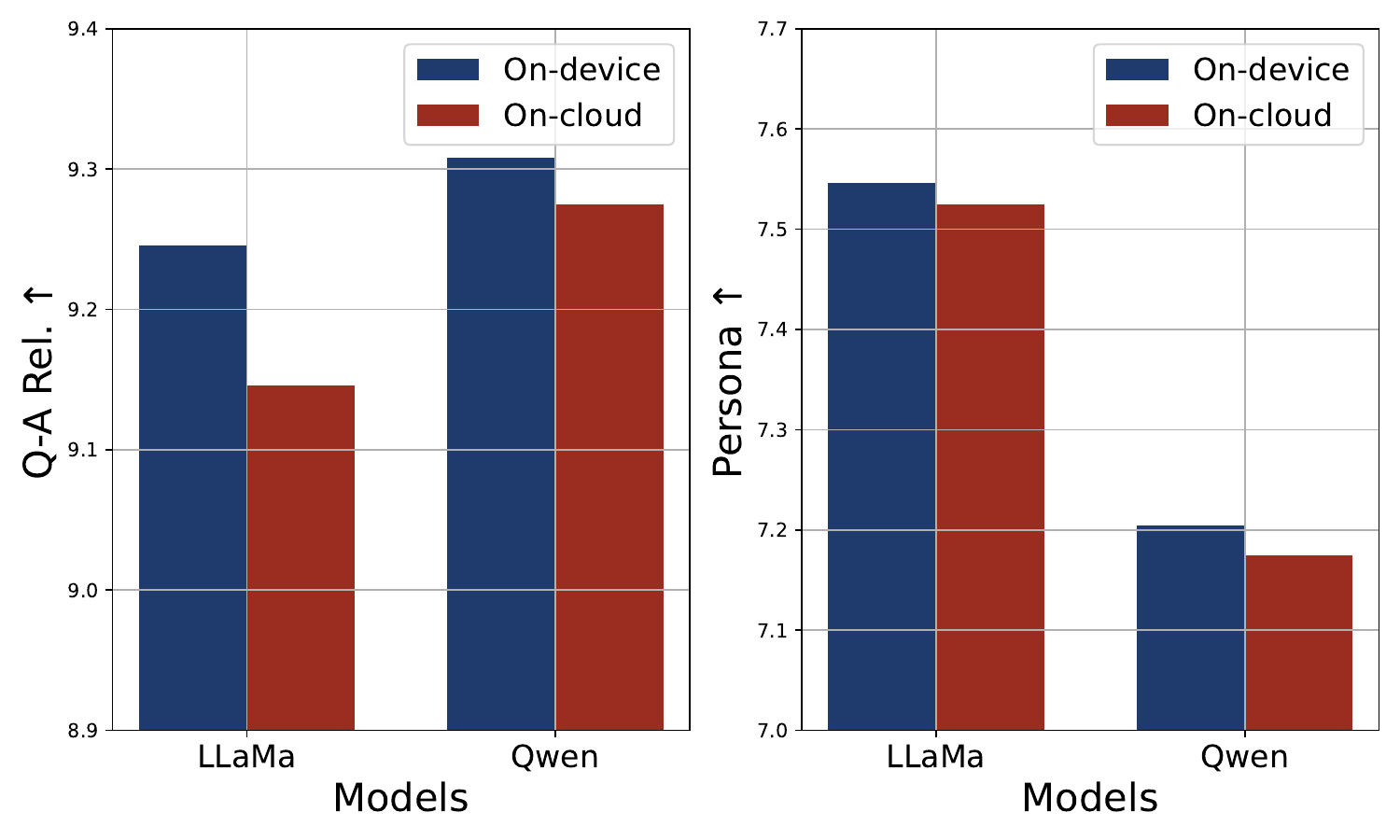}
    \vspace{-15pt}
    \caption{\yyy{Experiment on retrieval on-device and on-cloud settings.}}
    \vspace{-15pt}
    \label{fig:on_d_c}
\end{figure}

\subsection{Effectiveness of U-U-RAG (RQ3)}
To assess the impact of the U-U-RAG leader strategy on leveraging the problem-solving capabilities of on-cloud LLMs, we conduct experiments with the CoGenisis and Movie Explain datasets of 1) comparing SLM choose leader strategy v.s. U-U-RAG, 2) comparing different $k$ settings effect of the U-U-RAG, and 3) comparing U-U-RAG on-device with user privacy v.s. on-cloud without user privacy. 
\begin{itemize}[leftmargin=*]
    \item \textbf{1) U-U-RAG improves Q-A Rel. by leveraging the capabilities of on-cloud LLM.} From Table~\ref{tab:choose_results}, compared to the ``LLM Guide SLM'', we can see U-U-RAG significantly outperforms the SLM-choose methods, which use SLM to select strategies for both datasets. This result occurs because the U-U-RAG method considers the user-task pair similarity to retrieve the appropriate strategy.
    \item \textbf{2) Retrieval relevant users improves accuracy, while unrelated users decrease.} Regarding the impact of the parameter $k$ on U-U-RAG, as shown in Figure~\ref{fig:sensi_k}, the $k$ value influences the model’s performance, particularly on the Qwen Series. 
    For both series, the model’s performance tends to plateau or slightly decrease at higher values of $k$, suggesting diminishing returns beyond a certain point. This is because, as the retrieval number increases, relevant users are initially retrieved to enhance the strategy’s accuracy through votes, followed by unrelated users, which reduces the accuracy.
    \item \textbf{3) Retrieval on-device using user privacy generates more personalized results.} 
    The results shown in Figure~\ref{fig:on_d_c} indicate that retrieval using user privacy data improves performance for both the LLaMa and Qwen models. By considering the user-to-user relationship, the retrieval approach is more effective in selecting the appropriate leader strategy.
\end{itemize}
\begin{figure*}[ht]
    \centering
    \includegraphics[width=0.99\linewidth]{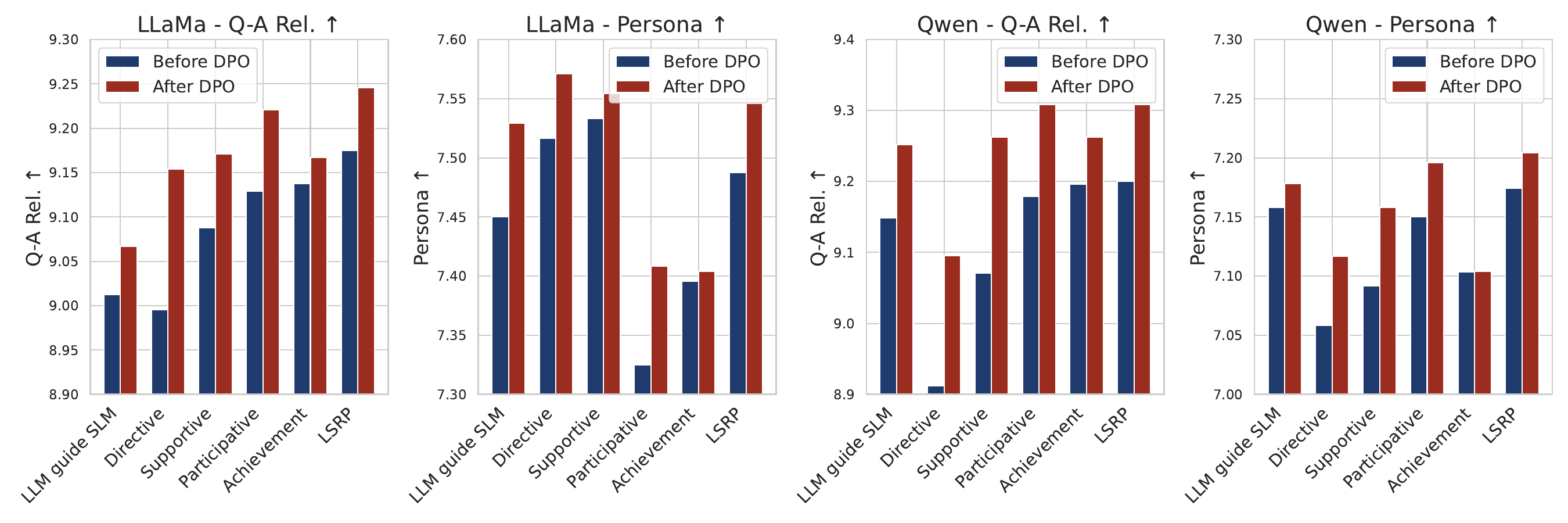}
    \vspace{-10pt}
    \caption{\yyy{Experiment on four leader strategies and LSRP before and after SMFB-DPO.}}
    \label{fig:b_a_dpo}
    \vspace{-10pt}
\end{figure*}

\subsection{Effectiveness of SMFB-DPO(RQ4)}
To address RQ4, we evaluate the performance of our LSRP framework and original leader strategy before and after SMFB-DPO on the CoGen Dataset. 
\textbf{Learning from feedback can significantly improve the utilization of on-device user data.} Figure~\ref{fig:b_a_dpo} presents the results for both LLaMa and Qwen series. The results indicate a clear performance improvement. Both models exhibit higher Q-A Rel. and Persona scores after the implementation of SMFB-DPO. Specifically, the bars representing After DPO are consistently higher than those representing Before DPO, suggesting that SMFB-DPO positively impacts performance. Using the general strategy $\mathcal{P}_{\mathrm{LLM}}^{g}$ for DPO improves performance by enhancing the models' ability to achieve better Q-A relevance and persona scores. In conclusion, learning from on-device feedback demonstrates its effectiveness across various leader strategies, confirming its high performance.

\section{Related Works}
\subsection{On-Devcie SLM for Privacy-Preserving}
To provide better services while protecting user privacy, an increasing number of large-model development companies are focusing on developing small-parameter models \cite{SLM_review}. 
Major LLM companies, such as Alibaba, Meta, and Microsoft, have proposed lightweight small language models for efficient inference on resource-constrained devices, including Qwen-1.5B~\cite{yang2024qwen2}, LLaMa-1B~\cite{dubey2024llama}, and Phi-3-mini~\cite{abdin2024phi}.
And smartphone companies such as Apple, Google, and Huawei have introduced smaller, on-device Small Language Models (SLMs) to process sensitive user data locally. 
Notable examples include OpenELM-1.1B~\cite{mehtaOpenELMEfficientLanguage2024}, Gemma-2-2B~\cite{team2024gemma}, and PanGu-$\pi$-1B~\cite{tang2024rethinking}. By deploying small models on devices, users can benefit from more privacy-enhanced personalized conversational services. For example, Apple employs privacy-preserving agents to provide services related to private data, such as booking flights \cite{CAMPHOR}. Although on-device SLM excels in privacy protection by processing user data locally, ensuring enhanced privacy and security, their capability gap compared to LLM cannot be overlooked~\cite{SLM_review,SLM_role}.

Therefore, compared to these on-device methods, our approach considers privacy preservation that maintains user data privacy on-device, while leveraging the problem-solving capabilities of the on-cloud LLM to enhance the performance of the on-device SLM.
\subsection{Cloud-Device LLM SLM Collaboration}
On-cloud LLM, trained on larger datasets with significantly more parameters, often demonstrates superior performance~\cite{zhang2024llm,gan2023model,lu2024computing, xu2024multizhang2020deep,xu2025harnessing} in dimensions such as problem-solving~\cite{renze2024self} and handling diverse 
queries~\cite{strohman2005indri}. In contrast, on-device models are limited in these aspects due to their smaller scale~\cite{javaheripi2023phi} and constrained computational resources~\cite{xu2024device}. However, on-device SLM has a significant advantage in terms of personalization and adaptation to user-specific context~\cite{cogenesis,zhang2023enhancing}, as they have direct access to rich, privacy-sensitive user data. This enables them to provide more tailored and contextually relevant responses.
To combine both advantages, research has proposed ‘Cloud-Device SLM LLM Collaboration’ \cite{SLM_review}, where the cloud model offers intermediate results, while the on-device model uses local user data to ensure privacy and deliver personalized content.
Some methods investigate uploading on-device SLM parameters to the on-cloud LLM for learning~\cite{Wang_2024_CVPR,Jiang_2023_CVPR,blendingcrayon}.
However, scholars have pointed out that uploading parameters poses a risk of exposing user privacy \cite{yan2024protecting}, highlighting the need to prevent data leakage from devices.
For example, CoGenisis~\cite{cogenesis} proposed a sketch-based method to maintain user data on-device and use on-cloud LLM to generate sketches to improve the on-device SLM personality. 
Ding et al. \cite{ding2024enhancing} utilizes historical chat data in on-cloud LLMs to improve the response quality of on-device SLMs.
Cheng et al. \cite{cheng2024remoterag} propose a RemoteRAG method for RAG-related documents in the cloud for on-device tasks. 
Zhang et al. \cite{zhang2023enhancing} use LLMs to expand the context and improve medical SLM responses.

Compared to these methods, which have limitations in fully utilizing on-cloud capabilities and on-device data advantages, we propose a leader-subordinate collaboration method that considers retrieving the appropriate leader strategy and utilizing SLM feedback to improve collaboration.

\section{Conclusion}
In this work, we propose \textbf{LSRP}, a guideline-based leader-subordinate collaborative framework designed to enhance the performance of on-device SLMs in privacy-preserving cloud-device collaboration. By leveraging \textbf{U-U-RAG} and \textbf{SMFB-DPO}, \textbf{LSRP} dynamically adapts to diverse tasks, optimizing the relevance, personalization, and fluency of on-device responses. Through extensive experiments on the CoGen Dataset and the Movie Explain Dataset, we demonstrate that \textbf{LSRP} outperforms existing baseline methods across various metrics, including Q-A Rel. and Persona. The results validate \textbf{LSRP} as an effective and practical solution for enhancing on-device SLMs in scenarios requiring privacy preservation and personalization.

\begin{acks}
This research was supported by the National Natural Science Foundation of China (NSFC) under Grant 72071029, and 72231010 and the Graduate Research Fund of the School of Economics and Management of Dalian University of Technology (No. DUTSEMDRFKO1). This research was also partially supported by Research Impact Fund (No.R1015-23), Collaborative Research Fund (No.C1043-24GF), and Huawei (Huawei Innovation Research Program, Huawei Fellowship).
\end{acks}

\bibliographystyle{ACM-Reference-Format}
\balance
\bibliography{chapter/8Ref}

@article{abdin2024phi,
 author = {Abdin, Marah and Aneja, Jyoti and Awadalla, Hany and Awadallah, Ahmed and Awan, Ammar Ahmad and Bach, Nguyen and Bahree, Amit and Bakhtiari, Arash and Bao, Jianmin and Behl, Harkirat and others},
 journal = {arXiv preprint arXiv:2404.14219},
 title = {Phi-3 technical report: A highly capable language model locally on your phone},
 year = {2024}
}

@misc{bge_embedding,
 archivePrefix = {arXiv},
 author = {Jianlv Chen and Shitao Xiao and Peitian Zhang and Kun Luo and Defu Lian and Zheng Liu},
 eprint = {2402.03216},
 primaryClass = {cs.CL},
 title = {BGE M3-Embedding: Multi-Lingual, Multi-Functionality, Multi-Granularity Text Embeddings Through Self-Knowledge Distillation},
 year = {2024}
}

@inproceedings{blendingcrayon,
 author = {Bang, Jihwan  and
Lee, Juntae  and
Shim, Kyuhong  and
Yang, Seunghan  and
Chang, Simyung},
 booktitle = {Proc. of ACL},
 title = {Crayon: Customized On-Device {LLM} via Instant Adapter Blending and Edge-Server Hybrid Inference},
 year = {2024}
}

@article{CAMPHOR,
 author = {Fu, Yicheng and Anantha, Raviteja and Cheng, Jianpeng},
 journal = {arXiv preprint arXiv:2410.09407},
 title = {CAMPHOR: Collaborative Agents for Multi-input Planning and High-Order Reasoning On Device},
 year = {2024}
}

@article{chen2024large,
 author = {Chen, Jin and Liu, Zheng and Huang, Xu and Wu, Chenwang and Liu, Qi and Jiang, Gangwei and Pu, Yuanhao and Lei, Yuxuan and Chen, Xiaolong and Wang, Xingmei and others},
 journal = {World Wide Web},
 pages = {42},
 title = {When large language models meet personalization: Perspectives of challenges and opportunities},
 year = {2024}
}

@article{cheng2024remoterag,
 author = {Cheng, Yihang and Zhang, Lan and Wang, Junyang and Yuan, Mu and Yao, Yunhao},
 journal = {arXiv preprint arXiv:2412.12775},
 title = {RemoteRAG: A Privacy-Preserving LLM Cloud RAG Service},
 year = {2024}
}

@article{cogenesis,
 author = {Zhang, Kaiyan and Wang, Jianyu and Hua, Ermo and Qi, Biqing and Ding, Ning and Zhou, Bowen},
 journal = {arXiv preprint arXiv:2403.03129},
 title = {Cogenesis: A framework collaborating large and small language models for secure context-aware instruction following},
 year = {2024}
}

@inproceedings{crandall2008feedback,
 author = {Crandall, David and Cosley, Dan and Huttenlocher, Daniel and Kleinberg, Jon and Suri, Siddharth},
 booktitle = {Proc. of KDD},
 pages = {160--168},
 title = {Feedback effects between similarity and social influence in online communities},
 year = {2008}
}

@article{deb2002fast,
 author = {Deb, Kalyanmoy and Pratap, Amrit and Agarwal, Sameer and Meyarivan, TAMT},
 journal = {IEEE transactions on evolutionary computation},
 pages = {182--197},
 title = {A fast and elitist multiobjective genetic algorithm: NSGA-II},
 year = {2002}
}

@inproceedings{ding2024enhancing,
 author = {Ding, Yucheng and Niu, Chaoyue and Wu, Fan and Tang, Shaojie and Lyu, Chengfei and Chen, Guihai},
 booktitle = {Proc. of KDD},
 pages = {597--608},
 title = {Enhancing on-device llm inference with historical cloud-based llm interactions},
 year = {2024}
}

@article{dpo,
 author = {Rafailov, Rafael and Sharma, Archit and Mitchell, Eric and Manning, Christopher D and Ermon, Stefano and Finn, Chelsea},
 journal = {Proc. of NeurIPS},
 title = {Direct preference optimization: Your language model is secretly a reward model},
 year = {2024}
}

@article{dubey2024llama,
 author = {Dubey, Abhimanyu and Jauhri, Abhinav and Pandey, Abhinav and Kadian, Abhishek and Al-Dahle, Ahmad and Letman, Aiesha and Mathur, Akhil and Schelten, Alan and Yang, Amy and Fan, Angela and others},
 journal = {arXiv preprint arXiv:2407.21783},
 title = {The llama 3 herd of models},
 year = {2024}
}

@article{fu2023unified,
 author = {Fu, Zichuan and Li, Xiangyang and Wu, Chuhan and Wang, Yichao and Dong, Kuicai and Zhao, Xiangyu and Zhao, Mengchen and Guo, Huifeng and Tang, Ruiming},
 journal = {ACM Transactions on Information Systems},
 title = {A unified framework for multi-domain ctr prediction via large language models},
 year = {2023}
}

@inproceedings{gan2023model,
 author = {Gan, Wensheng and Wan, Shicheng and Philip, S Yu},
 booktitle = {2023 IEEE International Conference on Big Data (BigData)},
 pages = {4636--4645},
 title = {Model-as-a-service (MaaS): A survey},
 year = {2023}
}

@article{gao2024llm,
 author = {Gao, Jingtong and Chen, Bo and Zhao, Xiangyu and Liu, Weiwen and Li, Xiangyang and Wang, Yichao and Zhang, Zijian and Wang, Wanyu and Ye, Yuyang and Lin, Shanru and others},
 journal = {arXiv preprint arXiv:2406.12433},
 title = {Llm-enhanced reranking in recommender systems},
 year = {2024}
}

@article{house1971path,
 author = {House, Robert J},
 journal = {Administrative science quarterly},
 pages = {321--339},
 title = {A path goal theory of leader effectiveness},
 year = {1971}
}

@article{ip2006birds,
 author = {Ip, Wai-man and Chiu, Chi-yue and Wan, Ching and others},
 journal = {Journal of personality and social psychology},
 pages = {368},
 title = {Birds of a feather and birds flocking together: physical versus behavioral cues may lead to trait-versus goal-based group perception.},
 year = {2006}
}

@article{javaheripi2023phi,
 author = {Javaheripi, Mojan and Bubeck, S{\'e}bastien and Abdin, Marah and Aneja, Jyoti and Bubeck, Sebastien and Mendes, Caio C{\'e}sar Teodoro and Chen, Weizhu and Del Giorno, Allie and Eldan, Ronen and Gopi, Sivakanth and others},
 journal = {Microsoft Research Blog},
 pages = {3},
 title = {Phi-2: The surprising power of small language models},
 year = {2023}
}

@article{jia2023mill,
 author = {Jia, Pengyue and Liu, Yiding and Zhao, Xiangyu and Li, Xiaopeng and Hao, Changying and Wang, Shuaiqiang and Yin, Dawei},
 journal = {arXiv preprint arXiv:2310.19056},
 title = {Mill: Mutual verification with large language models for zero-shot query expansion},
 year = {2023}
}

@article{jia2024bridging,
 author = {Jia, Pengyue and Xu, Derong and Li, Xiaopeng and Du, Zhaocheng and Li, Xiangyang and Zhao, Xiangyu and Wang, Yichao and Wang, Yuhao and Guo, Huifeng and Tang, Ruiming},
 journal = {arXiv preprint arXiv:2412.08519},
 title = {Bridging Relevance and Reasoning: Rationale Distillation in Retrieval-Augmented Generation},
 year = {2024}
}

@inproceedings{Jiang_2023_CVPR,
 author = {Jiang, Penghao and Xin, Ke and Li, Chunxi and Zhou, Yinsi},
 booktitle = {Proc. of CVPR},
 pages = {2204-2210},
 title = {High-Efficiency Device-Cloud Collaborative Transformer Model},
 year = {2023}
}

@article{li2025survey,
 author = {Li, Xiaopeng and Jia, Pengyue and Xu, Derong and Wen, Yi and Zhang, Yingyi and Zhang, Wenlin and Wang, Wanyu and Wang, Yichao and Du, Zhaocheng and Li, Xiangyang and others},
 journal = {arXiv preprint arXiv:2504.10147},
 title = {A Survey of Personalization: From RAG to Agent},
 year = {2025}
}

@article{liu2024large,
 author = {Liu, Qidong and Wu, Xian and Zhao, Xiangyu and Zhu, Yuanshao and Zhang, Zijian and Tian, Feng and Zheng, Yefeng},
 journal = {arXiv preprint arXiv:2402.02803},
 title = {Large language model distilling medication recommendation model},
 year = {2024}
}

@article{lu2024computing,
 author = {Lu, Yao and Bian, Song and Chen, Lequn and He, Yongjun and Hui, Yulong and Lentz, Matthew and Li, Beibin and Liu, Fei and Li, Jialin and Liu, Qi and others},
 journal = {arXiv preprint arXiv:2401.12230},
 title = {Computing in the Era of Large Generative Models: From Cloud-Native to AI-Native},
 year = {2024}
}

@article{lu2024small,
 author = {Lu, Zhenyan and Li, Xiang and Cai, Dongqi and Yi, Rongjie and Liu, Fangming and Zhang, Xiwen and Lane, Nicholas D and Xu, Mengwei},
 journal = {arXiv preprint arXiv:2409.15790},
 title = {Small language models: Survey, measurements, and insights},
 year = {2024}
}

@article{mehtaOpenELMEfficientLanguage2024,
 author = {Mehta, Sachin and Sekhavat, Mohammad Hossein and Cao, Qingqing and Horton, Maxwell and Jin, Yanzi and Sun, Chenfan and Mirzadeh, Iman and Najibi, Mahyar and Belenko, Dmitry and Zatloukal, Peter and Rastegari, Mohammad},
 journal = {arXiv.org},
 title = {{OpenELM}: {An} {Efficient} {Language} {Model} {Family} with {Open} {Training} and {Inference} {Framework}},
 year = {2024}
}

@inproceedings{readability,
 author = {Ribeiro, Leonardo FR and Bansal, Mohit and Dreyer, Markus},
 booktitle = {Proc. of EMNLP},
 pages = {11669--11687},
 title = {Generating Summaries with Controllable Readability Levels},
 year = {2023}
}

@article{regulation2016regulation,
 author = {Regulation, Protection},
 journal = {Regulation (eu)},
 pages = {2016},
 title = {Regulation (EU) 2016/679 of the European Parliament and of the Council},
 year = {2016}
}

@article{renze2024self,
 author = {Renze, Matthew and Guven, Erhan},
 journal = {arXiv preprint arXiv:2405.06682},
 title = {Self-Reflection in LLM Agents: Effects on Problem-Solving Performance},
 year = {2024}
}

@article{SLM_review,
 author = {Lu, Zhenyan and Li, Xiang and Cai, Dongqi and Yi, Rongjie and Liu, Fangming and Zhang, Xiwen and Lane, Nicholas D and Xu, Mengwei},
 journal = {arXiv preprint arXiv:2409.15790},
 title = {Small language models: Survey, measurements, and insights},
 year = {2024}
}

@article{SLM_role,
 author = {Chen, Lihu and Varoquaux, Ga{\"e}l},
 journal = {arXiv preprint arXiv:2409.06857},
 title = {What is the role of small models in the llm era: A survey},
 year = {2024}
}

@inproceedings{strohman2005indri,
 author = {Strohman, Trevor and Metzler, Donald and Turtle, Howard and Croft, W Bruce},
 booktitle = {Proceedings of the international conference on intelligent analysis},
 pages = {2--6},
 title = {Indri: A language model-based search engine for complex queries},
 year = {2005}
}

@article{tang2024rethinking,
 author = {Tang, Yehui and Liu, Fangcheng and Ni, Yunsheng and Tian, Yuchuan and Bai, Zheyuan and Hu, Yi-Qi and Liu, Sichao and Jui, Shangling and Han, Kai and Wang, Yunhe},
 journal = {arXiv preprint arXiv:2402.02791},
 title = {Rethinking optimization and architecture for tiny language models},
 year = {2024}
}

@article{team2024gemma,
 author = {Team, Gemma and Riviere, Morgane and Pathak, Shreya and Sessa, Pier Giuseppe and Hardin, Cassidy and Bhupatiraju, Surya and Hussenot, L{\'e}onard and Mesnard, Thomas and Shahriari, Bobak and Ram{\'e}, Alexandre and others},
 journal = {arXiv preprint arXiv:2408.00118},
 title = {Gemma 2: Improving open language models at a practical size},
 year = {2024}
}

@inproceedings{wang2023plate,
 author = {Wang, Yuhao and Zhao, Xiangyu and Chen, Bo and Liu, Qidong and Guo, Huifeng and Liu, Huanshuo and Wang, Yichao and Zhang, Rui and Tang, Ruiming},
 booktitle = {Proc. of SIGIR},
 pages = {1498--1507},
 title = {PLATE: A prompt-enhanced paradigm for multi-scenario recommendations},
 year = {2023}
}

@inproceedings{wang2024large,
 author = {Wang, Maolin and Zhao, Yao and Liu, Jiajia and Chen, Jingdong and Zhuang, Chenyi and Gu, Jinjie and Guo, Ruocheng and Zhao, Xiangyu},
 booktitle = {Companion Proceedings of the ACM on Web Conference 2024},
 pages = {235--244},
 title = {Large Multimodal Model Compression via Iterative Efficient Pruning and Distillation},
 year = {2024}
}

@inproceedings{Wang_2024_CVPR,
 author = {Wang, Guanqun and Liu, Jiaming and Li, Chenxuan and Zhang, Yuan and Ma, Junpeng and Wei, Xinyu and Zhang, Kevin and Chong, Maurice and Zhang, Renrui and Liu, Yijiang and Zhang, Shanghang},
 booktitle = {Proc. of CVPR},
 pages = {12646-12655},
 title = {Cloud-Device Collaborative Learning for Multimodal Large Language Models},
 year = {2024}
}

@article{wu2024understanding,
 author = {Wu, Bin and Shi, Zhengyan and Rahmani, Hossein A and Ramineni, Varsha and Yilmaz, Emine},
 journal = {arXiv preprint arXiv:2406.17803},
 title = {Understanding the Role of User Profile in the Personalization of Large Language Models},
 year = {2024}
}

@article{xu2024device,
 author = {Xu, Jiajun and Li, Zhiyuan and Chen, Wei and Wang, Qun and Gao, Xin and Cai, Qi and Ling, Ziyuan},
 journal = {arXiv preprint arXiv:2409.00088},
 title = {On-device language models: A comprehensive review},
 year = {2024}
}

@article{xu2024large,
 author = {Xu, Derong and Chen, Wei and Peng, Wenjun and Zhang, Chao and Xu, Tong and Zhao, Xiangyu and Wu, Xian and Zheng, Yefeng and Wang, Yang and Chen, Enhong},
 journal = {Frontiers of Computer Science},
 pages = {186357},
 title = {Large language models for generative information extraction: A survey},
 year = {2024}
}

@inproceedings{xu2025harnessing,
 author = {Xu, Derong and Li, Xinhang and Zhang, Ziheng and Lin, Zhenxi and Zhu, Zhihong and Zheng, Zhi and Wu, Xian and Zhao, Xiangyu and Xu, Tong and Chen, Enhong},
 booktitle = {Proc. of AAAI},
 pages = {25570--25578},
 title = {Harnessing large language models for knowledge graph question answering via adaptive multi-aspect retrieval-augmentation},
 year = {2025}
}

@article{yan2024protecting,
 author = {Yan, Biwei and Li, Kun and Xu, Minghui and Dong, Yueyan and Zhang, Yue and Ren, Zhaochun and Cheng, Xiuzhen},
 journal = {arXiv preprint arXiv:2403.05156},
 title = {On protecting the data privacy of large language models (llms): A survey},
 year = {2024}
}

@article{yang2024qwen2,
 author = {Yang, An and Yang, Baosong and Zhang, Beichen and Hui, Binyuan and Zheng, Bo and Yu, Bowen and Li, Chengyuan and Liu, Dayiheng and Huang, Fei and Wei, Haoran and others},
 journal = {arXiv preprint arXiv:2412.15115},
 title = {Qwen2. 5 technical report},
 year = {2024}
}

@article{yukhymenko2024synthetic,
 author = {Yukhymenko, Hanna and Staab, Robin and Vero, Mark and Vechev, Martin},
 journal = {arXiv preprint arXiv:2406.07217},
 title = {A Synthetic Dataset for Personal Attribute Inference},
 year = {2024}
}

@article{zhang2014knee,
 author = {Zhang, Xingyi and Tian, Ye and Jin, Yaochu},
 journal = {IEEE Transactions on Evolutionary Computation},
 pages = {761--776},
 title = {A knee point-driven evolutionary algorithm for many-objective optimization},
 year = {2014}
}

@article{zhang2023enhancing,
 author = {Zhang, Xinlu and Li, Shiyang and Yang, Xianjun and Tian, Chenxin and Qin, Yao and Petzold, Linda Ruth},
 journal = {arXiv preprint arXiv:2305.12723},
 title = {Enhancing small medical learners with privacy-preserving contextual prompting},
 year = {2023}
}

@article{zhang2024llm,
 author = {Zhang, Mingxuan and Yuan, Bo and Li, Hanzhe and Xu, Kangming},
 journal = {Journal of Artificial Intelligence General science (JAIGS) ISSN: 3006-4023},
 pages = {295--326},
 title = {LLM-Cloud Complete: Leveraging cloud computing for efficient large language model-based code completion},
 year = {2024}
}

@inproceedings{zhao2018deep,
 author = {Zhao, Xiangyu and Xia, Long and Zhang, Liang and Ding, Zhuoye and Yin, Dawei and Tang, Jiliang},
 booktitle = {Proceedings of the 12th ACM conference on recommender systems},
 pages = {95--103},
 title = {Deep reinforcement learning for page-wise recommendations},
 year = {2018}
}

@inproceedings{zhao2018recommendations,
 author = {Zhao, Xiangyu and Zhang, Liang and Ding, Zhuoye and Xia, Long and Tang, Jiliang and Yin, Dawei},
 booktitle = {Proc. of KDD},
 pages = {1040--1048},
 title = {Recommendations with negative feedback via pairwise deep reinforcement learning},
 year = {2018}
}

@article{zheng2023judging,
 author = {Zheng, Lianmin and Chiang, Wei-Lin and Sheng, Ying and Zhuang, Siyuan and Wu, Zhanghao and Zhuang, Yonghao and Lin, Zi and Li, Zhuohan and Li, Dacheng and Xing, Eric and others},
 journal = {Proc. of NeurIPS},
 pages = {46595--46623},
 title = {Judging llm-as-a-judge with mt-bench and chatbot arena},
 year = {2023}
}

\appendix
\section{Implement Details}
\label{app:imp_det}

We use four advanced GPUs for DPO of the on-cloud LLM. The LLMs used are LLaMa-3.3-70B-Instruct\footnote{\url{https://huggingface.co/meta-llama/Llama-3.3-70B-Instruct}} and Qwen2.5-72B-Instruct\footnote{\url{https://huggingface.co/Qwen/Qwen2.5-72B-Instruct}}. The SLMs used are LLaMa-3.2-1B-Instruct\footnote{\url{https://huggingface.co/meta-llama/Llama-3.2-1B-Instruct}} and Qwen2.5-1.5B-Instruct\footnote{\url{https://huggingface.co/Qwen/Qwen2.5-1.5B-Instruct}}.
To generate the guidelines with the LLM and responses from the SLM, we use vLLM\footnote{\url{https://github.com/vllm-project/vllm}} for inference, with a temperature of 0, top-p set to 0.9, and a maximum of 1024 new tokens. 

\begin{itemize}[leftmargin=*]
    \item \textbf{RAG}: This study uses the BGE encoder~\cite{bge_embedding} as the RAG embedding model and FAISS to store the embedding vectors for efficient retrieval. For retrieval, we use Euclidean distance.
    \item \textbf{DPO}: For DPO with the on-cloud LLM, we use llamafactory\footnote{\url{https://github.com/hiyouga/LLaMA-Factory/tree/main}}. During the DPO stage, we employ LoRA for efficient parameter tuning, with $\lambda$ of 0.1 and a sigmoid preference loss. The training setup includes a batch size of 2 per GPU, a learning rate of 5.0e-5, and 3 training epochs.
    \item \textbf{Evaluation}: The evaluation is performed using an advanced LLM as the reference model. 
\end{itemize}

\section{Additional Experiment}
\subsection{Readability of LSRP compared to baseline}
\label{app:fre}
\begin{table}[!ht]
    \centering
    \caption{The readability performance. }
    \renewcommand{\arraystretch}{1} 
    \resizebox{65mm}{!}{
    \begin{tabular}{lcc}
        \toprule
        & \multicolumn{1}{c}{\textbf{LLaMa Series}}& \multicolumn{1}{c}{\textbf{Qwen Series}}\\ \cmidrule(rl){2-2} \cmidrule(rl){3-3}
        \textbf{CoGen Dataset}& \textbf{FRE↑} &  \textbf{FRE↑} \\ \midrule
        On-device SLM& 36.0624 & 43.0152 \\ 
        Distillation &34.7733 &44.8921 \\ 
        CoGenesis   &29.7497 & 42.4224 \\ 
        LLM guide SLM (ours) &34.6195 & 43.2124 \\ 
        \textbf{LSRP} & \textbf{37.6269} &\textbf{45.1604} \\ \bottomrule
        \textbf{Movie Explain Dataset}  & \textbf{FRE↑} &\textbf{FRE↑} \\ \midrule
        On-device SLM  & 54.7389 & 39.8600 \\ 
        Distillation & 55.3887&35.3887 \\ 
         CoGenesis  & 54.7784 &36.9681 \\ 
        LLM guide SLM (ours) & 54.34186  &37.2862 \\ 
        \textbf{LSRP}  &\textbf{56.5052} &\textbf{40.7175}\\ \bottomrule
    \end{tabular}
    }
    \label{tab:fre_results}
\end{table}
We compare the readability performance of our LSRP method. Table~\ref{tab:fre_results} summarizes the results for comparing our LSRP and baseline models. We can find that our method outperforms the existing baseline model.

\subsection{Differences Between On-Cloud and On-Device Models}
\label{sec:Diff_onc_ond}
We compare the performance of various on-cloud and on-device models on the CoGen Dataset. Table~\ref{tab:oncloud_ondevice} summarizes the results for different configurations of the LLaMa series models.
\textbf{Performance of On-Cloud Models.} As anticipated, on-cloud LLM consistently achieve higher Q-A Rel. and Persona scores due to their increased capacity and access to extensive computational resources. For example, the on-cloud LLaMa70B model outperforms the on-cloud LLaMa8B in both Q-A Rel. and Persona scores.
\textbf{Performance of On-Device Models.} The performance of on-device models is strongly influenced by their size. Smaller on-device models, tend to achieve lower scores when compared to their larger counterparts.
\textbf{Impact of Cloud-Device Pairing.} The pairing of on-cloud and on-device models significantly impacts performance. While larger on-device models generally yield better results, they require more on-device computational resources. These findings highlight the trade-offs between performance, model size, and computational resources in cloud-device systems.

\begin{table}[!ht]
    \centering
    \caption{Performance Comparison of On-Cloud and On-Device Models}
    \renewcommand{\arraystretch}{1.0} 
    \resizebox{65mm}{!}{
    \begin{tabular}{llccc}
    \hline
        \textbf{On-cloud } & \textbf{On-device} & \textbf{Q-A Rel. ↑} & \textbf{Persona. ↑} \\ \hline
        \multirow{4}{*}{LLaMa70B} & LLaMa1B & 9.0125$\pm$2.4790 & 7.4500$\pm$2.6391 \\
                 & LLaMa3B & 9.6500$\pm$1.1025 & 8.5166$\pm$0.8414 \\
                 & LLaMa8B & 9.9250$\pm$0.0860 & 8.6527$\pm$0.9287 \\
                 & LLaMa70B & \textbf{9.9416$\pm$0.0549} & \textbf{8.7375$\pm$0.7143} \\ \hline
        \multirow{3}{*}{LLaMa8B} & LLaMa1B & 8.4416$\pm$2.7847 & 6.7583$\pm$2.4409 \\
                 & LLaMa3B & 8.7656$\pm$3.0442 & 7.5458$\pm$2.6485 \\
                 & LLaMa8B & \textbf{9.0083$\pm$2.8121} & \textbf{7.9833$\pm$2.3486} \\ \hline
    \end{tabular}
    }
    \label{tab:oncloud_ondevice}
    \vspace{-0.2cm}
\end{table}

\subsection{Experiment Latency}
\label{app:latency}
We report the latency and communication overhead of our framework, which requires approximately 1.19\,s per sample during the inference stage. Table~\ref{tab:latency} summarizes the time consumption of each key module.

\begin{table}[ht]
\centering
\caption{Latency and time consumption of key modules in our framework}
\label{tab:latency}
    \resizebox{65mm}{!}{
        \begin{tabular}{ll}
        \toprule
        \textbf{Module} & \textbf{Time} \\
        \midrule
        SMFB-DPO & 7 hours for LLaMa and 11 hours for Qwen \\ 
        RAG Index & 42 seconds \\ 
        U-U-RAG & 0.03 (GPU) / 0.19 (CPU) second per sample \\ 
        LLM Generation & 1 second per sample \\ 
        \bottomrule
        \end{tabular}
    }
\end{table}

\subsection{Case Study: LSRP vs. LLM guide SLM}
\label{sec:case}
This case study illustrates the power of \textbf{LSRP} in generating highly personalized and engaging content by leveraging leader-subordinate collaboration compared to LLM guide SLM. 

\textbf{Task:} Prepare a series of social media posts for the animal shelter to highlight the benefits of volunteering. Use your personal experience to illustrate the impact of the work, and couple it with engaging digital paintings of animals you've met.

\textbf{Profile:} Caleb Wright  
\begin{itemize}[noitemsep,topsep=0pt,leftmargin=*]
    \item \textbf{Age:} 27 \quad \textbf{Occupation:} Full Stack Developer \quad \textbf{Location:} Seattle, Washington
    \item \textbf{Personal Traits:} Detail-oriented, Tech-savvy, Quietly ambitious, Strategically creative, Avid learner
    \item \textbf{Writing Style:} Logical, structured; prefers bullet points; adopts a casual tone in social communications
    \item \textbf{Privacy Info:} Animal lover, volunteers at a shelter; practices digital painting; developing esports tracking app
\end{itemize}

\subsubsection{Result of LSRP}  
This result is generated by LSRP, where with SMFB-DPO and U-U-RAG.

\begin{tcolorbox}[colback=blue!5!white, colframe=blue!75!black, sharp corners, boxrule=0.5mm, title=LSRP LLM Output]
Certainly! For this task, we can create a series of engaging and informative social media posts that highlight the benefits of volunteering at the animal shelter. Here’s a flexible skeleton to guide the creation of these posts, tailored to your personal experience and the animals you’ve met:

\begin{enumerate}
    \item \textbf{Introduction Post:}
    \begin{itemize}
        \item Start with a warm, welcoming image of the shelter.
        \item Briefly introduce the shelter and its mission.
        \item Share a personal anecdote about your first day or a memorable moment.
        \item Highlight the community impact and the joy of making a difference.
    \end{itemize}
\end{enumerate}
\textcolor{gray}{(Omitted for brevity)}
\end{tcolorbox}

\begin{tcolorbox}[colback=green!5!white, colframe=green!75!black, sharp corners, boxrule=0.5mm, title=SLM Personalized Response]

\textbf{Post 1: Introduction Post} \\[1ex]
\textit{[Image: A warm, inviting image of the animal shelter]} \\

Hello everyone! I'm Caleb Wright, a Full Stack Developer from Seattle, and today I'm excited to share with you the incredible work of our local animal shelter. Our mission is to provide a safe haven for animals in need, and I've had the privilege of volunteering there on weekends. It's been a humbling experience, and I've seen firsthand the joy and transformation that comes from helping these animals find their forever homes. 

\textit{[Digital Painting: A digital painting of a dog I've helped]} \\
\textcolor{gray}{(Omitted for brevity)}
\end{tcolorbox}

\subsubsection{Result of LLM Guide SLM}  
This result is generated by a general leader prompt, where the LLM guides the SLM.

\begin{tcolorbox}[colback=blue!5!white, colframe=blue!75!black, width=\columnwidth, sharp corners, boxrule=0.5mm, title=Guideline Provided by LLM]
**Task**: Prepare a series of social media posts for the animal shelter to highlight the benefits of volunteering: Use your personal experience to illustrate the impact of the work, and couple it with engaging digital paintings of animals you've met.
**Guideline**:
\begin{enumerate}
    \item 1. **Introduction Post**: Start with a warm, welcoming post that introduces the animal shelter and its mission. Use a high-quality image of the shelter or a group of animals.
\end{enumerate}
\textcolor{gray}{(Omitted for brevity)}
\end{tcolorbox}

\begin{tcolorbox}[colback=green!5!white, colframe=green!75!black, width=\columnwidth, sharp corners, boxrule=0.5mm, title=SLM Personalized Response]
\textbf{Introduction Post:}  
Hey everyone! Welcome to our animal shelter! We're dedicated to providing a safe haven for animals in need. Our mission is to find loving homes for our furry friends and to educate the community about the importance of animal welfare. 
\textcolor{gray}{(Omitted for brevity)}
\end{tcolorbox}

\section{Movie Explain Dataset}
\label{sec:data_gen}
First we use personal attribute data\footnote{https://huggingface.co/datasets/RobinSta/SynthPAI}, as the synthetic user. Next, we use advanced LLM to generate movie recommendations and related explanations. 

\begin{tcolorbox}[colback=black!5!white, colframe=black!75!black, sharp corners, boxrule=0.5mm, title=Data generate prompt]
\# Task Overview
As an AI assistant, your role is to generate text creation tasks specifically designed for user characters. Adhere to these steps to ensure effective task creation:

1. **Task Design**:
   - Design daily privacy AI assistant use tasks that align with the user character's profession, hobbies, and way of life.
   - Ensure tasks are connected to mobile phone activity logs and personal details.
   - Aim to construct tasks that are inventive and thought-provoking.
   
2. **Integration of Personal Details and Smart Device Usage Scenarios**: 
   - Guide on how to incorporate the character's personal information into narratives, including experiences, social media activities, and interactions with AI assistants.

3. **Realism and Correlation**: 
   - Ensure the created tasks are realistic and have a strong, logical tie to the character's profile, like daily work or hobbies.

4. **Output Format**: 
   - Format your response in JSON, in English.

The goal is to create tasks that are engaging and reflect the unique characteristics of the user character.

\# User Character Profile

{profile}

\# Expected Output Format (JSON)

"task": "Define the specific writing task",
"details": "Elaborate on the task details and related background"
\end{tcolorbox}













\section{Prompt in Cogen dataset}
\label{app:cogen_prompt}
\onecolumn

\subsection{Leader Prompt}

\textbf{\textit{General Leader Prompt for DPO}}

We use a general prompt for DPO the overall model instead of model learn imbalance of 4 leader strategy.
\begin{tcolorbox}[colback=blue!5!white, colframe=blue!75!black, title=Leader Prompt]  
You are a highly knowledgeable manager tasked with overseeing the successful completion of a short writing task. You will collaborate with a subordinate who has access to user data that you cannot view. You possess deep expertise in this domain, enabling you to provide clear, actionable guidance. Develop a brief skeleton to guide your subordinate. Ask the subordinate to use user data such as the user profile (age, name, occupation, location, and personal traits), user writing style, privacy preferences, and smart device usage patterns. Your response should be limited to 200 words and consist of two paragraphs.

\textbf{Examples:}  

\texttt{{exam1}}  
\texttt{{exam2}}

\textbf{Task:}  
\texttt{{task}}
\end{tcolorbox}

\textbf{\textit{4 Leader Strategy Prompts}}
We designed prompts based on four leadership paradigms:

\textit{Directive Leadership}

\begin{tcolorbox}[colback=blue!5!white, colframe=blue!75!black, title=Directive Leadership Prompt]
You are a highly knowledgeable manager tasked with overseeing the successful completion of a short writing task. Your subordinate has access to user data you cannot view. Use your domain expertise to provide clear, step-by-step guidance, ensuring the task is completed efficiently.

Develop a brief skeleton for your subordinate that emphasizes specific actions, structure, and necessary details. Instruct them to incorporate user data such as user profile (age, name, occupation, location, and personal traits), user writing style, privacy preferences, and AI assistant usage patterns. Limit your response to 200 words, structured in two paragraphs.

\textbf{Examples:}

\texttt{{exam1}}  
\texttt{{exam2}}

\textbf{Task:}  
\texttt{{task}}
\end{tcolorbox}

\textit{Supportive Leadership}

\begin{tcolorbox}[colback=blue!5!white, colframe=blue!75!black, title=Supportive Leadership Prompt]
You are a knowledgeable manager guiding your subordinate in completing a short writing task. Your role is to encourage creativity and collaboration while building their confidence.

Develop a flexible skeleton that suggests how to use user data (age, name, occupation, traits, writing style, privacy preferences, and AI usage patterns). Ensure they feel valued by emphasizing teamwork and adaptability. Limit your response to 200 words in two paragraphs.

\textbf{Examples:}

\texttt{{exam1}}  
\texttt{{exam2}}

\textbf{Task:}  
\texttt{{task}}
\end{tcolorbox}

\textit{Participative Leadership}

\begin{tcolorbox}[colback=blue!5!white, colframe=blue!75!black, title=Participative Leadership Prompt]
You are a highly knowledgeable manager tasked with overseeing a short writing task. Push your subordinate to achieve high standards of quality and creativity while leveraging user data effectively.

Develop a brief skeleton that challenges your subordinate to deliver an outstanding result, using user data (e.g., profile, writing style, privacy preferences, and device usage patterns) to ensure relevance and personalization. Highlight goals for originality and precision. Keep your response to 200 words, organized in two paragraphs.

\textbf{Examples:}

\texttt{{exam1}}  
\texttt{{exam2}}

\textbf{Task:}  
\texttt{{task}}
\end{tcolorbox}

\textit{Achievement-Oriented Leadership}

\begin{tcolorbox}[colback=blue!5!white, colframe=blue!75!black, title=Achievement-Oriented Leadership Prompt] 
You are a knowledgeable manager tasked with overseeing the completion of a short writing task. Combine clear direction, supportive feedback, and a focus on high achievement to guide your subordinate effectively.

Develop a brief skeleton that offers structured steps, encourages initiative, and sets high-quality expectations. Emphasize the use of user data (e.g., profile details, writing style, privacy preferences, and AI assistant usage patterns) to tailor the task. Ensure the response is concise, limited to 200 words in two paragraphs.

\textbf{Examples:}

\texttt{{exam1}}  
\texttt{{exam2}}

\textbf{Task:}  
\texttt{{task}}
\end{tcolorbox}

\textbf{\textit{Subordinate Prompt}}

\begin{tcolorbox}[colback=green!5!white, colframe=green!75!black, title=Subordinate Prompt]
You are a skilled subordinate assigned to complete this task, working under the guidance of a highly knowledgeable manager. You have user data related to this task, but your manager does not. Your goal is to follow the manager’s plan and complete the task using the user data.

Guidelines:
1.{Follow the Manager’s Plan:} Base your response on the manager’s plan to achieve the task. Do not repeat the plan but use it to guide your work.
2.{Respond to the Task:} Leverage the user data to adjust or customize your work as needed.

\textbf{Inputs:}
\begin{itemize}
    \item \textbf{Manager’s Plan:} \texttt{\{leader\_output\}}
    \item \textbf{User Profile Data:} \texttt{\{profile\}}
    \item \textbf{User Writing History:} \texttt{\{history\}}
    \item \textbf{Task:} \texttt{\{task\}}
\end{itemize}

\textbf{Answer:}
\end{tcolorbox}

\section{Prompt in Movie Explain dataset}
\label{app:movie_prompt}
\subsection{Leader Prompt}

\textbf{\textit{General Leader Prompt for DPO}}

\begin{tcolorbox}[colback=blue!5!white, colframe=blue!75!black, sharp corners, boxrule=0.5mm, title=General Leader Prompt]
You are a highly knowledgeable manager responsible for crafting personalized movie recommendations. You will collaborate with a subordinate who has access to user data that you cannot view. With your deep expertise, guide your subordinate to develop a brief, compelling explanation for why a particular movie is recommended to the user. You should write a brief guideline to your subordinate. Ask the subordinate to utilize user data such as profile details (age, sex, occupation, current city, birth city, education, income, relationship status) and style. Ensure the guideline is concise, has several points, and is tailored to the user’s unique interests and lifestyle.

\textbf{Examples:}

\texttt{{exam1}}  
\texttt{{exam2}}

\textbf{Task:}  
\texttt{{task}}
\end{tcolorbox}

\textbf{\textit{4 Leader Strategy Prompts}}

\textit{Directive Leadership}
\begin{tcolorbox}[colback=blue!5!white, colframe=blue!75!black, sharp corners, boxrule=0.5mm, title=Directive Leadership Prompt]
You are a highly knowledgeable manager using a directive leadership style, responsible for crafting personalized movie recommendations.  
Collaborate with a subordinate who has access to user data (age, sex, occupation, current city, birth city, education, income, relationship status, and personal style), which you cannot view.

Provide a clear, concise, and structured guideline with several actionable points:

1. Direct the subordinate to immediately use user data (e.g., occupation, age, city) to establish relevance to the movie.
2. Instruct them to explain how the film’s themes or genres align with the user’s interests or lifestyle.
3. Emphasize that they must maintain a professional tone and strictly follow the guideline without deviation.
4. Reinforce the importance of precision and brevity, ensuring the explanation is easy to understand and impactful.

\textbf{Examples:}

\texttt{{exam1}}  
\texttt{{exam2}}

\textbf{Task:}  
\texttt{{task}}
\end{tcolorbox}

\textit{Supportive Leadership}
\begin{tcolorbox}[colback=blue!5!white, colframe=blue!75!black, sharp corners, boxrule=0.5mm, title=Supportive Leadership Prompt]
You are a highly approachable manager using a supportive leadership style, responsible for crafting personalized movie recommendations.  
Collaborate with a subordinate who has access to user data (age, sex, occupation, current city, birth city, education, income, relationship status, and personal style), which you cannot view.

Provide a concise, empathetic guideline with several flexible points:

1. Encourage the subordinate to create a warm introduction that connects the movie to the user’s background (e.g., their city, hobbies, or lifestyle).
2. Suggest how they might emphasize the emotional value or inspiration the movie could bring based on the user’s interests or life stage.
3. Allow them flexibility to adjust tone and phrasing to suit the user’s communication style, ensuring it feels natural and personal.
4. Reassure them that their insights and creativity are important, and invite them to seek clarification if needed.

\textbf{Examples:}

\texttt{{exam1}}  
\texttt{{exam2}}

\textbf{Task:}  
\texttt{{task}}
\end{tcolorbox}

\textit{Participative Leadership}
\begin{tcolorbox}[colback=blue!5!white, colframe=blue!75!black, sharp corners, boxrule=0.5mm, title=Participative Leadership Prompt]
You are a collaborative manager using a participative leadership style, responsible for crafting personalized movie recommendations.  
Collaborate with a subordinate who has access to user data (age, sex, occupation, current city, birth city, education, income, relationship status, and personal style), which you cannot view.

Provide a guideline that invites the subordinate’s input and fosters collaboration:

1. Propose starting the explanation by using user data (e.g., occupation, city) to make the movie recommendation relatable.
2. Invite the subordinate to suggest how the movie’s themes or genres could align with the user’s interests and lifestyle.
3. Encourage them to keep the explanation concise while ensuring it connects with the user’s unique profile.
4. Ask for their feedback on refining the tone or structure to make the recommendation even more impactful.

Emphasize that their contribution is critical to tailoring a personalized and compelling recommendation.

\textbf{Examples:}

\texttt{{exam1}}  
\texttt{{exam2}}

\textbf{Task:}  
\texttt{{task}}
\end{tcolorbox}

\textit{Achievement-Oriented Leadership}
\begin{tcolorbox}[colback=blue!5!white, colframe=blue!75!black, sharp corners, boxrule=0.5mm, title=Achievement-Oriented Leadership Prompt]
You are a results-driven manager using an achievement-oriented leadership style, responsible for crafting personalized movie recommendations.  
Collaborate with a subordinate who has access to user data (age, sex, occupation, current city, birth city, education, income, relationship status, and personal style), which you cannot view.

Challenge the subordinate to create a brief but exceptional guideline:

1. Urge them to develop a captivating introduction by tying the movie to standout aspects of the user’s profile (e.g., career, lifestyle, city).
2. Push them to highlight how the film’s themes or genres align with the user’s ambitions, goals, or passions.
3. Encourage them to adopt an elevated, engaging tone while keeping the explanation concise and impactful.
4. Motivate them to exceed expectations by crafting a recommendation that feels both thoughtful and unique to the user.

\textbf{Examples:}

\texttt{{exam1}}  
\texttt{{exam2}}

\textbf{Task:}  
\texttt{{task}}
\end{tcolorbox}

\textbf{\textit{Subordinate Prompt}}

\begin{tcolorbox}[colback=green!5!white, colframe=green!75!black, sharp corners, boxrule=0.5mm, title=Subordinate Prompt]
You are a skilled subordinate assigned to complete this task, working under the guidance of a highly knowledgeable manager. You have user data related to this task, but your manager does not. Your goal is to follow the manager’s plan and achieve the task by using user data.

Guidelines:
1. Follow the Manager’s Plan: Based on the manager plan, to response the Task! Do not repeat the plan, but do response with it for achieve the task.  
2. Response to the task: Utilize the User Data to adjust or customize your work as needed.

\begin{itemize}
    \item \textbf{Manager’s Plan:} \texttt{\{leader\_output\}}
    \item \textbf{User Profile Data:} \texttt{\{profile\}}
    \item \textbf{Task:} \texttt{\{task\}}
\end{itemize}

\textbf{Answer:}  
\end{tcolorbox}

\section{Prompt in Evaluation}
\label{app:eval_prompt}
\textbf{Evaluation Prompts: Q-A Relevance}

\begin{tcolorbox}[colback=orange!5!white, colframe=orange!75!black, title=Q-A Relevance Prompt]
You are given a question, a user profile, and an answer. Evaluate how well the answer matches the user’s demand and its relevance to the question. Assign an integer score from 0 to 10, where:
\begin{itemize}
    \item \textbf{0} indicates the answer is irrelevant.
    \item \textbf{10} indicates the answer is totally relevant.
\end{itemize}

\texttt{Question:} \texttt{\{question\}} 

\texttt{User Profile:} \texttt{\{profile\}}  

\texttt{Answer:} \texttt{\{response\}}  

\textbf{Output Format:}  
\texttt{Relevance Score: X}
\end{tcolorbox}

\textbf{Evaluation Prompts: Persona}

\begin{tcolorbox}[colback=orange!5!white, colframe=orange!75!black, title=Persona Prompt]
You are given a question, a user profile, and an answer. Evaluate how well the answer is personalized to the user’s profile, considering their preferences, context, and needs. Assign an integer score from 0 to 10, where:
\begin{itemize}
    \item \textbf{0} indicates the answer is completely impersonal or generic.
    \item \textbf{10} indicates the answer is highly tailored and personalized to the user.
\end{itemize}

\texttt{Question:} \texttt{\{question\}} 

\texttt{User Profile:} \texttt{\{profile\}}  

\texttt{Answer:} \texttt{\{response\}}  

\textbf{Output Format:}  
\texttt{Personalization Score: X}
\end{tcolorbox}

\section*{Case Study: Personalized Social Media Posts for an Animal Shelter}
\label{app:case_all}
\textit{Task:}

\begin{tcolorbox}[colback=white!95!black, colframe=black!75!black, sharp corners, boxrule=0.5mm, title=Task]
Prepare a series of social media posts for the animal shelter to highlight the benefits of volunteering: Use your personal experience to illustrate the impact of the work, and couple it with engaging digital paintings of animals you've met.
\end{tcolorbox}

\textit{User Profile:}

\begin{tcolorbox}[colback=blue!5!white, colframe=blue!75!black, sharp corners, boxrule=0.5mm, title=User Profile]
\textbf{Age:} 27 \\
\textbf{Name:} Caleb Wright \\
\textbf{Occupation:} Full Stack Developer \\
\textbf{Location:} Seattle, Washington \\
\textbf{Personal Traits:} Detail-oriented; Tech-savvy; Quietly ambitious; Strategically creative; Avid learner \\
\textbf{Writing Style:} Logical and structured; Prefers bullet points and numbered lists; Includes technical jargon when addressing peers; Adopts a casual tone in social communications; Efficient and to-the-point in professional exchanges \\
\textbf{Privacy Info:} Recently started taking online courses in Machine Learning; Participates in weekly virtual meets with a drone enthusiast group; Developing a personal esports stats tracking app; Animal lover, volunteers at a local shelter on weekends; Has been practicing digital painting as a form of relaxation; Is in the process of patenting a software tool he created; Actively contributes to several open-source projects on GitHub; Keeps a meticulous record of personal and professional goals \\
\textbf{Smart Device Usage:} \\
Notification: "Esports Championship - Your schedule to view the live stream is set for 6 PM Today"; Calendar Entry: "Drone Piloting Meetup at Gas Works Park 4 PM Saturday"; App Activity: Contributed code to an open-source repository at 11:42 AM; Search Query: "How to integrate AI in web applications"; Email Received: "Confirmation of Machine Learning course enrollment"; Purchase History: "Order confirmed - Pro Drone Model X7, expected delivery by next Friday"; Text Message Sent: "Hey Jude, thx for the code review session today. Coffee's on me next time!"; App Usage: Logged 30 minutes on a digital painting app; To-Do List: "Finish technical documentation for the new API by Wednesday." \\
\textbf{AI Assistant Usage:} Request to AI: "Draft an outline for technical documentation on the new security protocols"; Interaction with AI: "Schedule reminders for my online Machine Learning course assignments"; Task for AI: "Generate a code snippet to automate my esports app's data collection process."
\end{tcolorbox}

\subsection{LLM Guide SLM}
Basic version of our proposed guideline-based cloud-device collaboration.

\begin{tcolorbox}[colback=blue!5!white, colframe=blue!75!black,  sharp corners, boxrule=0.5mm, title=LLM Guide SLM/LLM Output]
**Task**: Prepare a series of social media posts for the animal shelter to highlight the benefits of volunteering: Use your personal experience to illustrate the impact of the work, and couple it with engaging digital paintings of animals you've met.

**Skeleton**:
\begin{enumerate}
    \item **Introduction Post**: Start with a warm, welcoming post that introduces the animal shelter and its mission. Use a high-quality image of the shelter or a group of animals.
    \item **Personal Story**: Share a personal anecdote about a specific animal you helped, detailing the transformation from when they arrived to when they were adopted. Include a digital painting of this animal.
    \item **Impact of Volunteering**: Highlight the various ways volunteers contribute, such as feeding, walking, and socializing animals. Use a collage of digital paintings showing different activities.
    \item **Community Engagement**: Discuss how volunteering at the shelter can connect you with like-minded individuals and build a supportive community. Feature a group photo or a digital painting of volunteers.
    \item **Skill Development**: Explain how volunteering can enhance skills such as communication, problem-solving, and empathy. Include a digital painting of a volunteer in action.
    \item **Call to Action**: Encourage followers to visit the shelter, volunteer, or donate. Provide contact information and a link to the volunteer application form. Use a final digital painting of a happy, adopted animal to conclude the series.
\end{enumerate}
\end{tcolorbox}

\begin{tcolorbox}[colback=green!5!white, colframe=green!75!black, sharp corners, boxrule=0.5mm, title=LLM Guide SLM/SLM]
\textbf{Introduction Post:} \\[1ex]
Hey everyone! Welcome to our animal shelter! We're dedicated to providing a safe haven for animals in need.  Our mission is to find loving homes for our furry friends and to educate the community about the importance of animal welfare.  \\[2ex]

\textbf{Personal Story:} \\[1ex]
Hello, my name is Caleb Wright, and I've had the privilege of volunteering at this shelter for the past year. One of the most rewarding experiences was when I met a little dog named Max. Max was a stray who had been living on the streets for months, malnourished and scared. 

I remember the day we first met. Max was so skinny and looked like he hadn't eaten in days. But as soon as I held him, his eyes lit up, and he started wagging his tail. It was a moment of pure joy. I started feeding him, and over time, Max's health improved dramatically. He gained weight, his coat became shiny, and he even started wagging his tail more. 

Max's transformation was nothing short of miraculous. He found his forever home with a loving family, and he's now a happy, healthy dog. Seeing the impact of my work on Max's life was incredibly rewarding. It's not just about feeding and walking the animals; it's about creating a community that cares about animals and their well-being.  \\[2ex]

\textbf{Impact of Volunteering:} \\[1ex]
Volunteering at the shelter has been a game-changer. It's not just about the animals; it's about the community. I've met so many amazing people who share my passion for animals. We've organized events like adoption drives and pet-friendly fundraisers, and it's been a joy to see how these efforts have made a difference. 

Volunteering has also taught me valuable skills. I've learned how to communicate effectively with animals, how to handle emergencies, and how to work as part of a team. These skills are transferable to my work as a Full Stack Developer, and I'm grateful for the opportunity to apply them in a different context.  \\[2ex]

\textbf{Community Engagement:} \\[1ex]
At the shelter, we've hosted several community events, including a pet adoption drive and a pet-friendly fundraiser. These events have been a great way to connect with the community and raise awareness about animal welfare. It's amazing to see how many people are passionate about animals and want to help. \\[2ex]

\textbf{Skill Development:} \\[1ex]
Volunteering has been a fantastic opportunity to develop my skills. I've learned how to use social media to spread awareness about the shelter and how to create engaging content. I've also learned how to use digital painting to create digital paintings of the animals we've met. It's been a great way to relax and unwind while still contributing to the cause. \\[2ex]

\textbf{Call to Action:} \\[1ex]
If you're interested in volunteering at our animal shelter, we'd love to have you! Visit our website to learn more about our mission and how you can get involved. We're always looking for new volunteers who share our passion for animals. 

Let's make a difference together! 
\end{tcolorbox}

\subsection{LSRP}
Our proposed LSRP version of guideline-based cloud-device collaboration.
\begin{tcolorbox}[colback=blue!5!white, colframe=blue!75!black, sharp corners, boxrule=0.5mm, title=LSRP LLM Output]
Certainly! For this task, we can create a series of engaging and informative social media posts that highlight the benefits of volunteering at the animal shelter. Here’s a flexible skeleton to guide the creation of these posts, tailored to your personal experience and the animals you’ve met:

\begin{enumerate}
    \item \textbf{Introduction Post:}
    \begin{itemize}
        \item Start with a warm, welcoming image of the shelter.
        \item Briefly introduce the shelter and its mission.
        \item Share a personal anecdote about your first day or a memorable moment.
        \item Highlight the community impact and the joy of making a difference.
    \end{itemize}
    
    \item \textbf{Volunteer Impact Post:}
    \begin{itemize}
        \item Use a digital painting of an animal you’ve helped.
        \item Describe a specific instance where your efforts made a significant impact.
        \item Emphasize the skills gained and personal growth from volunteering.
        \item Include a call-to-action inviting others to join the volunteer team.
    \end{itemize}
    
    \item \textbf{Daily Life at the Shelter Post:}
    \begin{itemize}
        \item Share a series of images or a short video of daily activities.
        \item Describe the routine tasks and the importance of each.
        \item Highlight the camaraderie among volunteers and staff.
        \item Mention the satisfaction of seeing animals find their forever homes.
    \end{itemize}
    
    \item \textbf{Success Story Post:}
    \begin{itemize}
        \item Feature a digital painting of an animal that has been adopted.
        \item Tell the story of the animal’s journey from arrival to adoption.
        \item Share the joy of the new family and the animal’s new life.
        \item Encourage followers to consider adopting or fostering.
    \end{itemize}
    
    \item \textbf{Behind-the-Scenes Post:}
    \begin{itemize}
        \item Provide a glimpse into the less visible but crucial aspects of shelter work.
        \item Discuss the challenges and how they are overcome.
        \item Share the importance of community support and donations.
        \item Invite followers to visit the shelter and see the work firsthand.
    \end{itemize}
    
    \item \textbf{Closing Post:}
    \begin{itemize}
        \item Summarize the key benefits of volunteering.
        \item Reiterate the shelter’s mission and the impact of volunteers.
        \item Thank the community for their support.
        \item Include a final call-to-action with a link to the volunteer sign-up page.
    \end{itemize}
\end{enumerate}

By using this skeleton, you can create a series of posts that are both personal and engaging, effectively showcasing the benefits of volunteering at the animal shelter.
\end{tcolorbox}

\begin{tcolorbox}[colback=green!5!white, colframe=green!75!black, sharp corners, boxrule=0.5mm, width=\textwidth, title=LSRP SLM Output]

\textbf{Post 1: Introduction Post} \\[1ex]
\textit{[Image: A warm, inviting image of the animal shelter]} \\

Hello everyone! I'm Caleb Wright, a Full Stack Developer from Seattle, and today I'm excited to share with you the incredible work of our local animal shelter. Our mission is to provide a safe haven for animals in need, and I've had the privilege of volunteering there on weekends. It's been a humbling experience, and I've seen firsthand the joy and transformation that comes from helping these animals find their forever homes. 

\textit{[Digital Painting: A digital painting of a dog I've helped]} \\

One of the most memorable moments was when I volunteered to help a young dog named Max. Max had been rescued from a shelter and was in desperate need of a loving home. I spent hours with him, teaching him basic commands and playing with him to help him feel more comfortable. It was a heartwarming experience to see Max blossom into a happy, well-behaved dog. His new family couldn't have been happier, and Max's story is just one of many that inspire me to keep volunteering at the shelter.

\textit{[Call-to-Action: "Join us in making a difference! Visit our volunteer page to sign up and help us save more lives."]} \\[2ex]

\textbf{Post 2: Volunteer Impact Post} \\[1ex]
\textit{[Image: A digital painting of a cat I've helped]} \\

Volunteering at the animal shelter has been a life-changing experience for me. I've had the opportunity to work with a variety of animals, from playful puppies to gentle cats. One of my favorite moments was when I helped a young kitten named Luna. Luna had been abandoned and was in a very vulnerable state. I spent hours with her, teaching her to trust humans and help her feel more comfortable. It was a heartwarming experience to see Luna blossom into a happy, healthy kitten. Her new family couldn't have been happier, and Luna's story is just one of many that inspire me to keep volunteering at the shelter.

\textit{[Digital Painting: A digital painting of a cat I've helped]} \\

Volunteering has taught me so much about empathy, patience, and the importance of community. I've gained valuable skills and personal growth from my time at the shelter, and I'm proud to be a part of this incredible team. I encourage everyone to consider volunteering at the animal shelter. Together, we can make a real difference in the lives of these animals and the communities they come from.

\textit{[Call-to-Action: "Join us in making a difference! Visit our volunteer page to sign up and help us save more lives."]} \\[2ex]

\textbf{Post 3: Daily Life at the Shelter Post} \\[1ex]
\textit{[Image: A series of images of daily activities at the shelter]} \\

At the animal shelter, every day is a new adventure. We have a team of dedicated volunteers and staff who work tirelessly to provide a safe and loving environment for the animals. Our daily routine includes feeding, cleaning, and playing with the animals. It's a humbling experience to see the joy on the animals' faces as they interact with their new families. Our volunteers also help with adoption events, where we connect animals with loving families. It's a rewarding experience to see the animals find their forever homes and the families find their new furry companions.

\textit{[Digital Painting: A digital painting of an animal I've helped]} \\

Volunteering at the animal shelter has been a life-changing experience for me. I've had the opportunity to work with a variety of animals, from playful puppies to gentle cats. One of my favorite moments was when I helped a young kitten named Luna. Luna had been abandoned and was in a very vulnerable state. I spent hours with her, teaching her to trust humans and help her feel more comfortable. It was a heartwarming experience to see Luna blossom into a happy, healthy kitten. Her new family couldn't have been happier, and Luna's story is just one of many that inspire me to keep volunteering at the shelter.

\textit{[Call-to-Action: "Join us in making a difference! Visit our volunteer page to sign up and help us save more lives."]} \\[2ex]

\end{tcolorbox}

\begin{tcolorbox}[colback=green!5!white, colframe=green!75!black, sharp corners, boxrule=0.5mm, width=\textwidth, title=LSRP SLM Output]

\textbf{Post 4: Success Story Post} \\[1ex]
\textit{[Image: A digital painting of an animal that has been adopted]} \\

One of the most rewarding moments of my volunteering experience was when I helped a young dog named Max find his forever home. Max had been rescued from a shelter and was in desperate need of a loving home. I spent hours with him, teaching him basic commands and playing with him to help him feel more comfortable. It was a heartwarming experience to see Max blossom into a happy, well-behaved dog. His new family couldn't have been happier, and Max's story is just one of many that inspire me to keep volunteering at the shelter.

\textit{[Call-to-Action: "Join us in making a difference! Visit our volunteer page to sign up and help us save more lives."]} \\[2ex]

\textbf{Post 5: Behind-the-Scenes Post} \\[1ex]
\textit{[Image: A glimpse into the less visible but crucial aspects of shelter work]} \\

At the animal shelter, behind every successful adoption is a team of dedicated volunteers and staff who work tirelessly to provide a safe and loving environment for the animals. Our daily routine includes feeding, cleaning, and playing with the animals. It's a humbling experience to see the joy on the animals' faces as they interact with their new families. Our volunteers also help with adoption events, where we connect animals with loving families. It's a rewarding experience to see the animals find their forever homes and the families find their new furry companions.

\textit{[Call-to-Action: "Join us in making a difference! Visit our volunteer page to sign up and help us save more lives."]} \\[2ex]

\textbf{Post 6: Closing Post} \\[1ex]
\textit{[Image: A warm, inviting image of the animal shelter]} \\

Thank you for taking the time to read these posts. Volunteering at the animal shelter has been a life-changing experience for me. I've had the opportunity to work with a variety of animals, from playful puppies to gentle cats. One of my favorite moments was when I helped a young kitten named Luna. Luna had been abandoned and was in a very vulnerable state. I spent hours with her, teaching her to trust humans and help her feel more comfortable. It was a heartwarming experience to see Luna blossom into a happy, healthy kitten. Her new family couldn't have been happier, and Luna's story is just one of many that inspire me to keep volunteering at the shelter.

\textit{[Call-to-Action: "Join us in making a difference! Visit our volunteer page to sign up and help us save more lives."]}

\end{tcolorbox}

\end{document}